\documentclass[12pt]{elsarticle}

\usepackage{listings}
\usepackage{graphicx}
\usepackage{color}
\usepackage{url}
\usepackage{paralist}
\usepackage{enumitem}
\usepackage{amsmath}
\usepackage{amsfonts}
\usepackage{amssymb}
\usepackage{proof}
\usepackage{placeins}
\usepackage{tikz}				%graphics package, uses PGF
\usetikzlibrary {shapes.geometric, arrows}  %circle and rectangle
\definecolor{commentcolor}{gray}{0.6}
\definecolor{codebackgroundcolor}{rgb}{0.97,0.99,0.94}
\definecolor{paragraphnumbercolor}{rgb}{0.99,0.60,0.60}
\definecolor{jpcolor}{rgb}{0.2,0.0,0.8}
\definecolor{bacolor}{rgb}{0.0,0.5,0.0}
\definecolor{reconciliationcolor}{rgb}{0.2,0.6,0.2}
\definecolor{notbacolor}{rgb}{0.5,0.0,0.0}

\lstdefinelanguage{bless}
  { morekeywords=[2]{ %AADL keywords
    aadlboolean,aadlinteger,aadlreal,aadlstring,
    abstract,access,and,annex,applies,binding,bus,
    calls,classifier,compute,connections,constant,data,delta,device,
    end,enumeration,event,extends,false,feature,features,flow,
    flows,group,implementation,in,inherit,inverse,is,
    list,memory, mod,mode,modes,none,not,of,or,
    out,package,parameter,path,port,private,process,processor,
    properties,property,prototypes,provides,public,range,reference,
    refined,rem,renames,requires,self,set,sink,source,subcomponents,
    subprogram,system,thread,to,true,type,units,virtual,with,
    stop,abort},
  morekeywords=[1]{ %BLESS keywords
      assert,declare,
    availability,while,forall,exists,that,are,numberof,dispatch,sticky,all,
    aperiodic,sporadic,timed,hybrid,xor,implies,iff,sum,product,
    now,variables,if,fi,skip,invariant,bound,cand,cor,null,timeout,states,
    persistent,shared,transitions,computation,delay,
    initial,complete,final,state,count,fresh,on,
    BLESS,Action,Assertion,pre,post,for,  %subBLESS,
    variant,record,array,do,until,boolean,tops,
    shared,spread,fetchadd,fetchand,fetchor,fetchxor,swap,
    def,integer,natural,rational,real,complex,time,
    Typed,Invariant,Precondition,Postcondition,
    otherwise,then,else,elsif,abs,
    updated,frozen,always,never,pause,error,propagations,behavior,
    use,types,events},
  sensitive=true,
  morecomment=[l]{--},
  columns=[r]fixed,
  basicstyle=\ttfamily\footnotesize,
  tabsize=2,
  otherkeywords={<<,>>,&,??,@,',[,],^,:,+=,\{**,**\},->,
    =>,?,!,+=>,:=,
  *!<,*!>,!<,!>, %-[,]->,
    ~>,(,)}
  }

\definecolor{commentcolor}{gray}{0.6}
\definecolor{codebackgroundcolor}{rgb}{0.98,0.99,0.95}

\lstdefinelanguage{slang} 
{
	frame=single,
	frameround=tttt,
	backgroundcolor= \color{codebackgroundcolor},	
  morecomment=[l]{--},
  columns=[r]fixed,
  basicstyle=\ttfamily\footnotesize,
  tabsize=2,
  otherkeywords={<<,>>,&,??,@,',[,],^,:,+=,\{,\},->,
    =>,?,!,+=>,:,=,
  *!<,*!>,!<,!>, %-[,]->,
    ~>,(,)}
  }

\makeatletter
\def\gr@implitem#1<#2> #3 {%
   \sbox\z@{\hskip\labelsep\grammarlabel{#2}{#3}}%
   \strut\@@par%
   \vskip-\parskip%
   \vskip-\baselineskip%
   \hrule\@height\z@\@depth\z@\relax%
   \item[\unhbox\z@]%
   \catcode`\<\active%
}
\makeatother

\def\doframeit#1{\vbox{%
  \hrule height\fboxrule
    \hbox{%
      \vrule width\fboxrule \kern\fboxsep
      \vbox{\kern\fboxsep #1\kern\fboxsep }%
      \kern\fboxsep \vrule width\fboxrule }%
    \hrule height\fboxrule }}

\def\frameit{\smallskip \advance \linewidth by -7.5pt \setbox0=\vbox \bgroup
\strut \ignorespaces }

\def\endframeit{\ifhmode \par \nointerlineskip \fi \egroup
\doframeit{\box0}}
\renewcommand{\prod}[2]{{#1}\times{#2}}

\renewcommand{\brack}[1]{\textsf{\lbrack}{#1}\textsf{\rbrack}}

\newcommand{\truesym}{\mbox{\it true}}

\newcommand{\compv}{t}      %% meta-variable for a component/thread id
\newcommand{\portv}{p}      %% meta-variable for a port variable id
\newcommand{\portsv}{A}     %% port variables of a component
\newcommand{\iportsv}{I}    %% infrastructure ports of a component
\newcommand{\cvarsv}{V}     %% local state of a component

\newcommand{\compids}{\mbox{\it ThreadIds}}
\newcommand{\portids}{\mbox{\it PortIds}}
\newcommand{\variables}{\mbox{\it Var}}
\newcommand{\aps}{\mbox{\it APS}}
\newcommand{\ips}{\mbox{\it IPS}}

\newcommand{\inports}[1]{\portsv^{i}_{#1}}
\newcommand{\outports}[1]{\portsv^{o}_{#1}}

\newcommand{\inportsc}{\inports{\compv}}
\newcommand{\outportsc}{\outports{\compv}}
\newcommand{\modsym}{{\cal M}} % model M
\newcommand{\valsym}{{\cal V}} % value domain
\newcommand{\featsym}{\cal F} % set of all features of a component Comp \in C
\newcommand{\propsym}{\mathit Prop} % set of all properties over a component Comp
\newcommand{\portkindsym}{\mbox{\it PortKind}}
\newcommand{\portdirectionsym}{\mbox{\it PortDirection}}
\newcommand{\isdataportsym}{\mbox{\it isDataPort}}

\newcommand{\isinportsym}{\mbox{\it isInPort}}
\newcommand{\isoutportsym}{\mbox{\it isOutPort}}

\newcommand{\inportssym}{\mbox{\it InPorts}}
\newcommand{\outportssym}{\mbox{\it OutPorts}}

\newcommand{\conndestssym}{\mbox{\it ConnDests}}

\newcommand{\datasym}{\mbox{\em data}}
\newcommand{\eventsym}{\mbox{\em event}}
\newcommand{\eventdatasym}{\mbox{\em eventdata}}

\newcommand{\insym}{\mbox{\em in}}
\newcommand{\outsym}{\mbox{\em out}}

\newcommand{\systemsym}{\mbox{\em system}}
\newcommand{\processsym}{\mbox{\em process}}
\newcommand{\threadsym}{\mbox{\em thread}}
\newcommand{\devicesym}{\mbox{\em device}}

\newcommand{\modelportkind}[1]{\modsym.\portkindsym({#1})}
\newcommand{\modelportdirection}[1]{\modsym.\portdirectionsym({#1})}
\newcommand{\modelisdataport}[1]{\modsym.\isdataportsym({#1})}

\newcommand{\modelisinport}[1]{\modsym.\isinportsym({#1})}

\newcommand{\modelisoutport}[1]{\modsym.\isoutportsym({#1})}

\newcommand{\modeloutports}[1]{\modsym.\outportssym[{#1}]}
\newcommand{\modelinports}[1]{\modsym.\inportssym[{#1}]}

\newcommand{\modelconndest}[1]{\modsym.\conndestssym({#1})}

\newcommand{\dispatchprotocolsym}{\mbox{\it DispatchProtocol}}
\newcommand{\periodicsym}{\mbox{\em Periodic}}
\newcommand{\sporadicsym}{\mbox{\em Sporadic}}

\newcommand{\modeldispatchprotocol}[1]{\modsym.\dispatchprotocolsym({#1})}

\newcommand{\iniports}[1]{\iportsv^{i}_{#1}}
\newcommand{\outiports}[1]{\iportsv^{o}_{#1}}

\newcommand{\iniportsc}{\iniports{\compv}}
\newcommand{\outiportsc}{\outiports{\compv}}

\newcommand{\mapupdate}[3]{{#1}\brack{#2\,\mapsto\,{#3}}}

\newcommand{\maplookup}[2]{{#1}({#2})}

\newcommand{\queuev}{q}

\newcommand{\queuemake}[1]{\langle {#1} \rangle}
\newcommand{\queueempty}{\queuemake{}}
\newcommand{\eventpresent}{*}
\newcommand{\queuedevent}{\queuemake{\eventpresent}}

\newcommand{\cvars}[1]{\cvarsv_{#1}}
\newcommand{\cvarsc}{\cvars{\compv}}

\newcommand{\aadlrule}[1]{\mbox{\it {#1}}}

\newcommand{\tinfer}[2]{\begin{array}[b]{c} {#2} \\ \hline
                          {#1} \end{array}}
\newcommand{\tinferlabt}[3]{\begin{array}{l}
                              {#1}:
                              \\[2pt]
                              \tinfer{#2}{#3}
                              \end{array}}

\newcommand{\compcont}[6]{\langle {#1}, {#2}, {#3}, {#4}, {#5}, {#6} \rangle}
\newcommand{\portlist}{ps}

\newcommand{\dsnotenabled}{\mbox{\em NotEnabled}}

\newcommand{\dsv}{S_t}
\newcommand{\dstimetriggeredsym}{\mbox{\em TimeTriggered}}
\newcommand{\dseventtriggeredsym}{\mbox{\em EventTriggered}}
\newcommand{\evalDirDispatchThread}[3]{{#1}\,\stackrel{\mbox{\tiny
      Dispatch Threads}}{\longrightarrow}\,{#3}}
\newcommand{\evalrecinprelsym}[1]{\stackrel{\mbox{\tiny Receive\_Input}}{\longrightarrow_{#1}}}
\newcommand{\evalrecinp}[4]{{#1}\, \stackrel{\mbox{\tiny Receive\_Input}}{\longrightarrow_{#2}}\,{#4}}
\newcommand{\evalsendoutp}[4]{{#1}\,\stackrel{\mbox{\tiny Send\_Output}}{\longrightarrow_{#2}}\,{#4}}
\newcommand{\cmove}[2]{{#1}\,\stackrel{\mbox{\tiny Move}}{\longrightarrow}\,{#2}}

\newcommand{\sysphasev}{\mbox{\em Phs}}
\newcommand{\sysphaseinit}{\mbox{\em Initializing}}
\newcommand{\sysphasecompute}{\mbox{\em Computing}}
  
\newcommand{\systhreadsv}{\mbox{\em Thrs}}

\newcommand{\sysschedv}{\mbox{\em Schs}}

\newcommand{\sysschedwait}{\mbox{\em WaitingForDispatch}}

\newcommand{\syscommv}{\mbox{\em Comms}}

\newcommand{\sysstate}[4]{\langle {#1}, {#2}, {#3}, {#4} \rangle}

\newcommand{\slangInline}[1]{\texttt{#1}}
\newcommand{\slangGetValue}[1]{\slangInline{getValue}}
\newcommand{\slangPutValue}[1]{\slangInline{putValue}}
\newcommand{\slangSendOutput}[1]{\slangInline{sendOutput}}
\newcommand{\slangReceiveInput}[1]{\slangInline{receiveInput}}

\newcommand{\aadlInline}[1]{\texttt{#1}}
\newcommand{\aadlGetCount}[1]{\aadlInline{Get\_Count}}
\newcommand{\aadlGetValue}[1]{\aadlInline{Get\_Value}}
\newcommand{\aadlNextValue}[1]{\aadlInline{Next\_Value}}

\newcommand{\aadlDispatchStatussym}{\aadlInline{Dispatch\_Status}}

\newcommand{\aadlReceiveInputsym}{\aadlInline{Receive\_Input}}
\newcommand{\aadlSendOutputsym}{\aadlInline{Send\_Output}}

\newcommand{\aadlBaseTypessym}{\aadlInline{Base\_\-Types}}
\newcommand{\aadlDataModelsym}{\aadlInline{Data\_\-Model}}
\newcommand{\aadlQueueSizesym}{\aadlInline{Queue\_\-Size}}
\newcommand{\aadlTimeStampsym}{\aadlInline{Time\_\-Stamp}}

\newcommand{\aadlComputeEP}{\aadlInline{Compute\_\-Entrypoint}}
\newcommand{\aadlInitEP}{\aadlInline{Initialize\_\-Entrypoint}}

\newcommand{\commentunresolved}[1]{}  %% omit unresolved comments
\usepgflibrary{shapes.arrows}

\newcommand{\threadportconcepts}[1]		% {scale}
	{
	\sffamily\footnotesize

	\begin{tikzpicture}[scale=#1, line width=3pt]	%%make TiKZ picture
	 
	\fill[fill=black!20!white] (0,0) rectangle (3,10);
	\fill[fill=black!20!white] (11,0) rectangle (14,10);
	
	\draw[color=green!60!black] (7,9.75) node {\textit{Infrastucture-Application Boundary}};
	
	\draw[line width=2pt] (5.5,2.25) rectangle (8.75,8.5);
	\draw (7,9.1) node {\textit{Application Code}};
	\draw (7,8.75) node { \textit{(Entry Points)} };

	\fill[fill=black!20!white] (5.75,7) rectangle (8.5,8.25);
	\draw (7.13,7.75) node[scale=0.8] {\texttt{Initialize\_Entrypoint}};
	\fill[fill=black!20!white] (5.75,5.5) rectangle (8.5,6.75);
	\draw (7.13,6.25) node[scale=0.8] {\texttt{Compute\_Entrypoint}};
	\fill[fill=black!20!white] (5.75,4) rectangle (8.5,5.25);
	\draw (7.13,4.4) node[scale=0.8] {\texttt{Finalize\_Entrypoint}};
	\fill[fill=black!20!white] (5.75,2.5) rectangle (8.5,3.75);
	\draw (7.3,3.5) node[scale=0.8] {\texttt{Activate\_}};
	\draw (7.3,3.25) node[scale=0.8] {\texttt{Deactivate\_}};
	\draw (7.13,3) node[scale=0.8] {\texttt{and Recover\_}};
	\draw (7.13,2.75) node[scale=0.8] {\texttt{Entrypoint(s)}};
	
	\draw[fill=green!50!white, line width=1pt] (5.5,1.0) rectangle (8.75,2);
	\draw (7,1.75) node[scale=0.7]  {Local};
	\draw (7,1.5) node[scale=0.7]  {Persistent};
	\draw (7,1.25) node[scale=0.7]  {Variables};

	\node at (3,8)[fill=blue!20!white, single arrow, minimum height = 4cm, minimum width = 1.75cm] {};
	\draw[line width=0.5pt] (1.25,7.75) rectangle (2.75,8.25);
	\draw[line width=0.5pt] (1.5,7.75) -- (1.5,8.25);
	\draw[line width=0.5pt] (1.75,7.75) -- (1.75,8.25);
	\draw[line width=0.5pt] (2,7.75) -- (2,8.25);
	\draw[line width=0.5pt] (2.5,7.75) -- (2.5,8.25);
	\draw (2.25,8) node{\ldots};
	\draw[line width=0.5pt] (3.25,7.75) rectangle (4.75,8.25);
	\draw[line width=0.5pt] (3.5,7.75) -- (3.5,8.25);
	\draw[line width=0.5pt] (3.75,7.75) -- (3.75,8.25);
	\draw[line width=0.5pt] (4,7.75) -- (4,8.25);
	\draw[line width=0.5pt] (4.5,7.75) -- (4.5,8.25);
	\draw (4.25,8) node{\ldots};

	\node at (11,8)[fill=blue!20!white, single arrow, minimum height = 4cm, minimum width = 1.75cm] {};
	\draw[line width=0.5pt] (10,7.75) rectangle (10.5,8.25);
	\draw[line width=0.5pt] (11.5,7.75) rectangle (12,8.25);

	\node at (3,6)[fill=blue!20!white, single arrow, minimum height = 4cm, minimum width = 1.75cm] {};
	\draw[line width=0.5pt] (1.25,5.75) rectangle (2.75,6.25);
	\draw[line width=0.5pt] (1.5,5.75) -- (1.5,6.25);
	\draw[line width=0.5pt] (1.75,5.75) -- (1.75,6.25);
	\draw[line width=0.5pt] (2,5.75) -- (2,6.25);
	\draw[line width=0.5pt] (2.5,5.75) -- (2.5,6.25);
	\draw (2.25,6) node{\ldots};
	\draw[line width=0.5pt] (3.25,5.75) rectangle (4.75,6.25);
	\draw[line width=0.5pt] (3.5,5.75) -- (3.5,6.25);
	\draw[line width=0.5pt] (3.75,5.75) -- (3.75,6.25);
	\draw[line width=0.5pt] (4,5.75) -- (4,6.25);
	\draw[line width=0.5pt] (4.5,5.75) -- (4.5,6.25);
	\draw (4.25,6) node{\ldots};

	\node at (11,6)[fill=blue!20!white, single arrow, minimum height = 4cm, minimum width = 1.75cm] {};
	\draw[line width=0.5pt] (10,5.75) rectangle (10.5,6.25);
	\draw[line width=0.5pt] (11.5,5.75) rectangle (12,6.25);

	\node at (3,4)[fill=blue!20!white, single arrow, minimum height = 4cm, minimum width = 1.75cm] {};
	\draw[line width=0.5pt] (2,3.75) rectangle (2.5,4.25);
	\draw[line width=0.5pt] (3.5,3.75) rectangle (4,4.25);

	\node at (11,4)[fill=blue!20!white, single arrow, minimum height = 4cm, minimum width = 1.75cm] {};
	\draw[line width=0.5pt] (10,3.75) rectangle (10.5,4.25);
	\draw[line width=0.5pt] (11.5,3.75) rectangle (12,4.25);

	\draw[dash=on 10pt off 8pt phase 9pt, color=green!70!black] (3,0.5) -- (11,0.5) -- (11,9.5) -- (3,9.5) -- (3,0.5);
	
	\draw (3,8.5) node[color=blue] {In Event Data Ports};
	\draw (3,6.5) node[color=blue] {In Event Ports};
	\draw (3,4.5) node[color=blue] {In Data Ports};
	\draw (11,8.5) node[color=blue] {Out Event Data Ports};
	\draw (11,6.5) node[color=blue] {Out Event Ports};
	\draw (11,4.5) node[color=blue] {Out Data Ports};
	
   \draw[red, line width=2mm, arrows= {->[slant=0.5]}] (1.5,3.25) .. controls (2,2.5) and (4,2.5) .. (4.5,3.25);
   \filldraw[fill=red!30!white, line width=1pt] (1.5,2) rectangle (4.5,2.5);
   \draw (3,2.25) node  {\texttt{Receive\_Input}};	
	
   \draw[red, line width=2mm, arrows= {->[slant=0.5]}] (9.5,3.25) .. controls (10,2.5) and (12,2.5) .. (12.5,3.25);
   \filldraw[fill=red!30!white, line width=1pt] (9.5,2) rectangle (12.5,2.5);
   \draw (11,2.25) node  {\texttt{Send\_Output}};	
	
   \filldraw[fill=red!30!white, line width=1pt] (9.5,1) rectangle (12.5,1.5);
   \draw (11,1.25) node[scale=0.9]  {\texttt{Dispatch\_Status}};	

  \draw (1.75,9.25) node[scale=0.7] {\textit{Input Infrastructure}};
  \draw (1.75,9) node[scale=0.7] {\textit{Port State (IPS)}};
  \draw (4.1,9.25) node[scale=0.7] {\textit{Input Application}};
  \draw (4.1,9) node[scale=0.7] {\textit{Port State (APS)}};
  \draw (12.25,9.25) node[scale=0.7] {\textit{Output Infrastructure}};
  \draw (12.25,9) node[scale=0.7] {\textit{Port State (IPS)}};
  \draw (9.8,9.25) node[scale=0.7] {\textit{Output Application}};
  \draw (9.9,9) node[scale=0.7] {\textit{Port State (APS)}};

   \draw[red, line width=3pt, arrows= {->[slant=0.5]}] (4.5,4.9) .. controls (5,4.5) and (6,4.5) .. (6.5,4.9);
   \filldraw[fill=red!30!white, line width=1pt] (4.5,5) rectangle (6.5,5.67);
   \draw (5.5,5.5) node[scale=0.9]  {\texttt{Get\_Value}};	
   \draw (5.5,5.2) node[scale=0.9]  {\texttt{Next\_Value}};	

   \filldraw[fill=red!30!white, line width=1pt] (4.5,6.75) rectangle (6.5,7.25);
   \draw (5.5,7) node  {\texttt{Get\_Count}};	
   \draw[line width=1pt, arrows={->}]  (4.75,7.75) -- (5,7.25);
   \draw[line width=1pt, arrows={->}]  (4.75,6.25) -- (5,6.75);

   \filldraw[fill=red!30!white, line width=1pt] (5,3.5) rectangle (6.5,4);
   \draw (5.75,3.75) node[scale=0.9]  {\texttt{Updated}};	
   \draw[line width=1pt, arrows={->}]  (4,4) -- (5,3.75);

   \draw[red, line width=3pt, arrows= {->[slant=0.5]}] (7.75,4.9) .. controls (8.25,4.5) and (9.25,4.5) .. (9.75,4.9);
   \filldraw[fill=red!30!white, line width=1pt] (7.75,5) rectangle (9.75,5.5);
   \draw (8.75,5.25) node  {\texttt{Put\_Value}};	

	\draw[color=black] (13.5,5) node[rotate=270] {Communication Infrastructure};
	\draw[color=black] (0.5,5) node[rotate=90] {Communication Infrastructure};
	\end{tikzpicture}

	}	%end of \threadportconcepts

\newcommand{\ourthreadportconcepts}[1]		% {scale}
	{\sffamily\footnotesize
	\begin{tikzpicture}[scale=#1, line width=3pt]	%%make TiKZ picture
	 
	\fill[fill=black!20!white] (0,0) rectangle (3,10);
	\fill[fill=black!20!white] (11,0) rectangle (14,10);
	\draw[color=black] (0.5,5) node[rotate=90] {Communication Infrastructure};
	\draw[color=black] (13.5,5) node[rotate=270] {Communication Infrastructure};
	
	\draw[color=green!60!black] (7,9.75) node {\textit{Infrastructure-Application Boundary}};
	
	\draw[line width=2pt] (5.5,2.25) rectangle (8.75,8.5);
	\draw (7,9.1) node {\textit{Application Code}};
	\draw (7,8.75) node { \textit{(Entry Points)} };

	\fill[fill=black!20!white] (5.75,7) rectangle (8.5,8.25);
	\draw (7.13,7.75) node[scale=0.8] {\texttt{Initialize\_Entrypoint}};
	\fill[fill=black!20!white] (5.75,5.5) rectangle (8.5,6.75);
	\draw (7.13,6.25) node[scale=0.8] {\texttt{Compute\_Entrypoint}};
	\draw (7.13,6) node[scale=0.8] {\texttt{(Dispatch\_Status)}};
	\fill[fill=black!20!white] (5.75,4) rectangle (8.5,5.25);
	\draw (7.13,4.7) node[scale=0.8] {\texttt{Finalize\_Entrypoint}};
	\draw[fill=green!50!white, line width=1pt] (5.5,1.0) rectangle (8.75,2);
	\draw (7,1.75) node[scale=0.7]  {Local};
	\draw (7,1.5) node[scale=0.7]  {Persistent};
	\draw (7,1.25) node[scale=0.7]  {Variables};
      
	\node at (3,8)[fill=blue!20!white, single arrow, minimum height = 4cm, minimum width = 1.75cm] {};
	\draw[line width=0.5pt] (1.25,7.75) rectangle (2.75,8.25);
	\draw[line width=0.5pt] (1.5,7.75) -- (1.5,8.25);
	\draw[line width=0.5pt] (1.75,7.75) -- (1.75,8.25);
	\draw[line width=0.5pt] (2,7.75) -- (2,8.25);
	\draw[line width=0.5pt] (2.5,7.75) -- (2.5,8.25);
	\draw (2.25,8) node{\ldots};
	\draw[line width=0.5pt] (3.25,7.75) rectangle (4.75,8.25);
	\draw[line width=0.5pt] (3.5,7.75) -- (3.5,8.25);
	\draw[line width=0.5pt] (3.75,7.75) -- (3.75,8.25);
	\draw[line width=0.5pt] (4,7.75) -- (4,8.25);
	\draw[line width=0.5pt] (4.5,7.75) -- (4.5,8.25);
	\draw (4.25,8) node{\ldots};

	\node at (11,8)[fill=blue!20!white, single arrow, minimum height = 4cm, minimum width = 1.75cm] {};
	\draw[line width=0.5pt] (10,7.75) rectangle (10.5,8.25);
	\draw[line width=0.5pt] (11.5,7.75) rectangle (12,8.25);

	\node at (3,6)[fill=blue!20!white, single arrow, minimum height = 4cm, minimum width = 1.75cm] {};
	\draw[line width=0.5pt] (1.25,5.75) rectangle (2.75,6.25);
	\draw[line width=0.5pt] (1.5,5.75) -- (1.5,6.25);
	\draw[line width=0.5pt] (1.75,5.75) -- (1.75,6.25);
	\draw[line width=0.5pt] (2,5.75) -- (2,6.25);
	\draw[line width=0.5pt] (2.5,5.75) -- (2.5,6.25);
	\draw (2.25,6) node{\ldots};
	\draw[line width=0.5pt] (3.25,5.75) rectangle (4.75,6.25);
	\draw[line width=0.5pt] (3.5,5.75) -- (3.5,6.25);
	\draw[line width=0.5pt] (3.75,5.75) -- (3.75,6.25);
	\draw[line width=0.5pt] (4,5.75) -- (4,6.25);
	\draw[line width=0.5pt] (4.5,5.75) -- (4.5,6.25);
	\draw (4.25,6) node{\ldots};

	\node at (11,6)[fill=blue!20!white, single arrow, minimum height = 4cm, minimum width = 1.75cm] {};
	\draw[line width=0.5pt] (10,5.75) rectangle (10.5,6.25);
	\draw[line width=0.5pt] (11.5,5.75) rectangle (12,6.25);

	\node at (3,4)[fill=blue!20!white, single arrow, minimum height = 4cm, minimum width = 1.75cm] {};
	\draw[line width=0.5pt] (2,3.75) rectangle (2.5,4.25);
	\draw[line width=0.5pt] (3.5,3.75) rectangle (4,4.25);

	\node at (11,4)[fill=blue!20!white, single arrow, minimum height = 4cm, minimum width = 1.75cm] {};
	\draw[line width=0.5pt] (10,3.75) rectangle (10.5,4.25);
	\draw[line width=0.5pt] (11.5,3.75) rectangle (12,4.25);

	\draw[dash=on 10pt off 8pt phase 9pt, color=green!70!black] (3,0.5) -- (11,0.5) -- (11,9.5) -- (3,9.5) -- (3,0.5);
	
	\draw (3,8.5) node[color=blue] {In Event Data Ports};
	\draw (3,6.5) node[color=blue] {In Event Ports};
	\draw (3,4.5) node[color=blue] {In Data Ports};
	\draw (11,8.5) node[color=blue] {Out Event Data Ports};
	\draw (11,6.5) node[color=blue] {Out Event Ports};
	\draw (11,4.5) node[color=blue] {Out Data Ports};
	
   \draw[red, line width=2mm, arrows= {->[slant=0.5]}] (1.5,3.25) .. controls (2,2.5) and (4,2.5) .. (4.5,3.25);
   \filldraw[fill=red!30!white, line width=1pt] (1.5,2) rectangle (4.5,2.5);
   \draw (3,2.25) node  {\texttt{Receive\_Input}};	
	
   \draw[red, line width=2mm, arrows= {->[slant=0.5]}] (9.5,3.25) .. controls (10,2.5) and (12,2.5) .. (12.5,3.25);
   \filldraw[fill=red!30!white, line width=1pt] (9.5,2) rectangle (12.5,2.5);
   \draw (11,2.25) node  {\texttt{Send\_Output}};	

  \draw (1.75,9.25) node[scale=0.7] {\textit{Input Infrastructure}};
  \draw (1.75,9) node[scale=0.7] {\textit{Port State (IPS)}};
  \draw (4,9.25) node[scale=0.7] {\textit{Input Application}};
  \draw (4,9) node[scale=0.7] {\textit{Port State (APS)}};
  \draw (12.25,9.25) node[scale=0.7] {\textit{Output Infrastructure}};
  \draw (12.25,9) node[scale=0.7] {\textit{Port State (IPS)}};
  \draw (10,9.25) node[scale=0.7] {\textit{Output Application}};
  \draw (10,9) node[scale=0.7] {\textit{Port State (APS)}};

   \filldraw[fill=red!30!white, line width=1pt] (9.5,1) rectangle (12.5,1.5);
   \draw (11,1.25) node  {\texttt{Create\_Timeout}};

	\end{tikzpicture}
	}	%end of \languageinclusion

\newcommand{\threadlifecycle}[1]		% {scale}
	{
	\sffamily\footnotesize

	\begin{tikzpicture}[scale=#1, line width=1pt, rounded corners=10pt]	%%make TiKZ picture

   \draw (1,1)[ line width = 2pt] circle [x radius=1, y radius=0.5];
   \draw(1,1.1) node {thread};
   \draw(1,0.8) node {halted};

   \draw[line width=1pt, arrows={->}]  (2,1) .. controls (3,2) .. (3.5,2) -- (4.5,1);
   \draw (3.2,2.2) node[scale=0.7] {\tt Initialize\_Entrypoint};
   \draw (3.2,2.5) node[scale=0.7] {Director invokes};

   \draw[fill = blue!20!white] (4.5,0) rectangle (6.5,2);
   \draw (5.5,1.5) node[scale=0.8] {performing};
   \draw (5.5,1.2) node[scale=0.8] {\it thread};
   \draw (5.5,0.9) node[scale=0.8] {\it initialization};

   \draw[fill =  black!10!white] (5.5,0.3) circle [x radius=1, y radius=0.3];
   \draw (5.5,0.3) node[scale=0.8] {\tt Send\_Output};

   \draw[line width=1pt, arrows={->}]  (6.5,1) -- (8,1);
   \draw (7.25,1.4) node[scale=0.7] {complete};
   \draw (7.25,1.2) node[scale=0.7] {\it initialization};

   \draw[fill = green!20!white] (9,1) circle [x radius=1, y radius=0.7];
  \draw (9,1.3) node {suspended};
   \draw (9,1) node {awaiting};
   \draw (9,0.7) node {dispatch};

   \draw[fill = blue!20!white] (6,3.5) rectangle (12,7);
   \draw (9.25,6.75) node {performing \it thread computation};

   \draw[line width=1pt, arrows={->}]  (9.75,1.5) .. controls (10.5,1.5) .. (11,2) -- (11,3.85) ;
   \draw (10.5,2.4) node[scale=0.7] {dispatch}; 
   \draw (10.3,2.2) node[scale=0.7] {\it computation}; 
   
   \draw[line width=1pt, arrows={->}]   (11.1,4.8) -- (11.1,5.5);
   \draw (10.4, 5.2) node[scale=0.7]  {accept};
   \draw (10.4, 5) node[scale=0.7]  {dispatch};

   \draw[line width=1pt, arrows={->}]   (8.5,4) -- (8.5,1.6);
   \draw (7.4,3.3) node[scale=0.7] {thread invokes};
   \draw (7.4,3.05) node[scale=0.7] {\tt Await\_Dispatch};
    \draw (7.9,2.4) node[scale=0.7] {complete};
   \draw (7.8,2.2) node[scale=0.7] {\textit{computation}};

   \draw[line width=1pt]   (12,4.3) .. controls (12.25,4.3) .. (12.5,3.1)  -- (12.5,2.6);
   \draw[line width=1pt, arrows={->}]   (12.5,2.6) .. controls (12.5,1.5) .. (12,1) -- (10,1);
   \draw (11,0.85) node[scale=0.7] {defer dispatch};

   \draw[fill = yellow] (10.6,4.3) circle [x radius=1.35, y radius=0.5];
  \draw (10.7,4.6) node[scale=0.8] {evaluate};
   \draw (10.6,4.2) node[scale=0.8] {\tt Dispatch\_Status};

    \draw (9.9,3.05) node[scale=0.7] {\tt Compute\_Entrypoint};
   \draw (10.1,3.28) node[scale=0.7] {Director invokes};

   \draw[fill = black!10!white] (10.6,6) circle [x radius=1.35, y radius=0.5];
   \draw (10.65,6) node[scale=0.8] {\tt Receive\_Input};

   \draw[line width=1pt, arrows={->}]   (9.25,6) -- (8.75,6);

   \draw[fill = yellow] (7.4,6) circle [x radius=1.35, y radius=0.5];
   \draw (7.4,6) node {execute};

   \draw[line width=1pt, arrows={->}]   (7.4,5.5) -- (7.4,4.8);

   \draw[fill =  black!10!white] (7.4,4.3) circle [x radius=1.35, y radius=0.5];
   \draw (7.4,4.3) node[scale=0.8] {\tt Send\_Output};

   \draw[fill = blue!20!white] (0,4) rectangle (2,6);
   \draw (1,5.4) node {performing};
   \draw (1,5) node {\it thread};
   \draw (1,4.6) node {\it finalize};

   \draw[line width=1pt, arrows={->}]   (8.25,1.4) -- (2,5);
   \draw (4,4.75) node[scale=0.7] {\tt Finalize\_Entrypoint};
   \draw (4,5) node[scale=0.7] {Director invokes};

   \draw[line width=1pt, arrows={->}]   (1,4) -- (1,1.5);
   
	\end{tikzpicture}

	}	%end of \threadlifecycle

\newcommand{\directorcycle}[1]		% {scale}
	{
	\sffamily\footnotesize

	\begin{tikzpicture}[scale=#1, line width=1pt, rounded corners=10pt]	%%make TiKZ picture

   \draw[line width = 2pt] (1.75,6.5) circle [x radius=1.25, y radius=0.5];
   \draw (1.75,6.5) node {\it off};

   \draw[line width=1pt, arrows={->}]   (1.75,6) -- (1.75,5);
   \draw (1.25,5.5) node[scale=0.7] {power on};

   \draw[fill = blue!20!white] (0,3.5) rectangle (3.5,5);
  \draw (1.75,4.45) node[scale=0.9] {invoke};
   \draw (1.75,4.1) node[scale=0.9] {\tt Initialize\_Entrypoint};
  \draw (1.75,3.75) node[scale=0.9] {(all threads)};
  \draw (0,4.6) -- (3.5,4.6);
  \draw (1.75,4.8) node {\it initialize};

   \draw[line width=1pt, arrows={->}]   (1.25,3.5) -- (1.25,2.5);
   \draw (2.25,3) node[scale=0.7] {threads initialized};

   \draw[fill = red!20!white] (0,1) rectangle (2.5,2.5);
   \draw (1.25,1.6) node {\tt Move};
   \draw (0,2.1) -- (2.5,2.1);
   \draw (1.25,2.3) node {\it move};

   \draw[line width=1pt, arrows={->}]   (2.5,1.75) -- (4.25,1.75);
   \draw (3.3,2.2) node[scale=0.7] {events \& data};
   \draw (3.3,2) node[scale=0.7] {moved};

   \draw[fill = green!20!white] (4.25,1) rectangle (6.75,2.5);
   \draw (5.5,1.85) node[scale=0.9] {find dispatchable};
   \draw (5.5,1.6) node[scale=0.9] {threads and};
   \draw (5.5,1.25) node[scale=0.9] {run timers};
   \draw (4.25,2.1) -- (6.75,2.1);
   \draw (5.5,2.3) node {\it dispatch};

   \draw[line width=1pt, arrows={->}]   (6.75,1.75) -- (8.5,1.75);
   \draw (7.75,2.2) node[scale=0.7] {dispatchable};
   \draw (7.75,2) node[scale=0.7] {threads};

   \draw[line width=1pt, arrows={->}]   (6,2.5) -- (6,3.5);
  \draw (7,3)  node  {stop(system)};

   \draw[fill = blue!20!white] (8.5,1) rectangle (12,2.5);
  \draw (10.25,1.95) node[scale=0.9] {invoke};
   \draw (10.25,1.6) node[scale=0.9] {\tt Compute\_Entrypoint};
  \draw (10.25,1.25) node[scale=0.9] {(dispatchable threads)};
   \draw (8.5,2.1) -- (12,2.1);
   \draw (10.25,2.3) node {\it compute};

   \draw[line width=1pt, arrows={->}]   (10.5,1) .. controls (10.5,0) .. (1.25,0) -- (1.25,1);
  \draw (5.5,0.25)  node[scale=0.7] {threads suspended};

   \draw[fill = blue!20!white] (5,3.5) rectangle (8.5,5);
  \draw (6.75,4.45) node[scale=0.9] {invoke};
   \draw (6.75,4.1) node[scale=0.9] {\tt Finalize\_Entrypoint};
  \draw (6.75,3.75) node[scale=0.9] {(all threads)};
  \draw (5,4.6) -- (8.5,4.6);
  \draw (6.75,4.8) node {\it finalize};

   \draw[line width=1pt, arrows={->}]   (7,5) .. controls (7,6) .. (6.5,6.5) -- (3,6.5);
  \draw (5.5,6.75)  node[scale=0.7] {threads halted};

	\end{tikzpicture}

	}	%end of \directorcycle

\DeclareMathSymbol{\Tau}{\mathalpha}{operators}{"54}
\DeclareMathOperator*{\off}{off}

\definecolor{wffcolor}{rgb}{0,0,0.9}
\definecolor{formulacolor}{rgb}{0,0.1,1}
\definecolor{variablecolor}{rgb}{1,0.2,0}
\definecolor{classcolor}{rgb}{0.7,0.3,0.8}
\definecolor{actioncolor}{rgb}{1,0.8,0.2}
\definecolor{assertioncolor}{rgb}{1.0,0.7,0.0}
\definecolor{blcolor}{rgb}{0.1,0,.8}
\definecolor{assertedactioncolor}{rgb}{0,1,0}
\definecolor{gray}{rgb}{0.2,0.2,0.2}

\newcommand{\msc}[1]{{\color{variablecolor}#1}}
\newcommand{\mm}[1]{{\color{gray}#1}}
\newcommand{\lstba}[1]

\newcommand{\now}{\color{red}\mathsf{now}}

\newcommand{\oomit}[1]{}

\newtheorem{definition}{Definition}

\makeatletter
\def\ps@pprintTitle{%
  \let\@oddhead\@empty
  \let\@evenhead\@empty
  \def\@oddfoot{\reset@font\hfil\thepage\hfil}%
  \let\@evenfoot\@oddfoot
}
\makeatother

\begin{document}

\begin{frontmatter}

%% Title, authors and addresses

%% use the tnoteref command within \title for footnotes;
%% use the tnotetext command for theassociated footnote;
%% use the fnref command within \author or \affiliation for footnotes;
%% use the fntext command for theassociated footnote;
%% use the corref command within \author for corresponding author footnotes;
%% use the cortext command for theassociated footnote;
%% use the ead command for the email address,
%% and the form \ead[url] for the home page:
%% \title{Title\tnoteref{label1}}
%% \tnotetext[label1]{}
%% \author{Name\corref{cor1}\fnref{label2}}
%% \ead{email address}
%% \ead[url]{home page}
%% \fntext[label2]{}
%% \cortext[cor1]{}
%% \affiliation{organization={},
%%            addressline={}, 
%%            city={},
%%            postcode={}, 
%%            state={},
%%            country={}}
%% \fntext[label3]{}

\title{Formalization of the AADL Run-Time Services with Time}

%% use optional labels to link authors explicitly to addresses:
%% \author[label1,label2]{}
%% \affiliation[label1]{organization={},
%%             addressline={},
%%             city={},
%%             postcode={},
%%             state={},
%%             country={}}
%%
%% \affiliation[label2]{organization={},
%%             addressline={},
%%             city={},
%%             postcode={},
%%             state={},
%%             country={}}

\author[label1]{Brian Larson}

\affiliation[label1]{organization={Multitude Corporation},%Department and Organization
           % addressline={}, 
            city={St Paul},
            postcode={}, 
            state={Minnesota},
            country={USA}}

\author[label2]{Ehsan Ahmad}

\affiliation[label2]{organization={College of Computing and Informatics, Saudi Electronic University},%Department and Organization
           % addressline={}, 
            city={Riyadh},
            %postcode={}, 
            %state={Minnesota},
            country={Saudi Arabia}}

\begin{abstract}
%The Architecture Analysis \& Design Language (AADL) is an industry standard modeling language distinguished by its emphasis on strong semantics for modeling real-time embedded systems. 
The Architecture Analysis \& Design Language (AADL) is an  architecture description language for design of cyber-physical systems--machines controlled by software. The AADL standard,  SAE International AS5506D, describes Run-Time Services (RTS) to be provided to execute AADL models in accordance with semantics defined by the standard. The RTS of primary concern are \emph{transport services} and \emph{timing services}.  Although, the study presented in 
\cite{ISOLA2022} sets a foundation for the formal semantics of AADL, but without modeling time.  This paper extends and simplifies this formalization using a modal logic defined by a Kripke structure,  to explicitly include time. The RTS defined in the AADL standard are also expanded to support reactive state-transition machines of the Behavior Specification annex standard language (BA) and its closely-related, formally-defined counterpart, the Behavior Language for Embedded Systems with Software (BLESS). An example of AADL RTS with time, implemented by the High Assurance Modeling and Rapid Engineering for Embedded Systems (HAMR) for state-transition machine behavior written in BLESS, is also presented.

\end{abstract}

%%Graphical abstract
%\begin{graphicalabstract}
%\includegraphics{grabs}
%\end{graphicalabstract}

%%Research highlights
% \begin{highlights}
% \item This paper presents an extension to the formal semantics of AADL RTS specified in \cite{ISOLA2022} to include time.
% \item A modal logic based on Kripke structure is defined to establish ``worlds'' representing instants of system operation using @ as a modal operator to access other worlds.
% \item The AADL RTS are defined in terms of this modal (temporal) logic to describe the progression during execution of variable values for both threads and RTS infrastructure.
% \item Additional RTS not included in the AADL standard, but necessary for state-machine implementation of threads are defined.
% \item HAMR, augmented by BLESS plugins, generates executable byte code that demonstrates the use of the additional RTS functions, including timestamp, timeout, and deferred dispatch.
% \end{highlights}

\begin{keyword}

AADL \sep BLESS \sep HAMR \sep cyber-physical systems \sep architecture \sep run-time services \sep temporal logic \sep Sireum
%% keywords here, in the form: keyword \sep keyword

%% PACS codes here, in the form: \PACS code \sep code

%% MSC codes here, in the form: \MSC code \sep code
%% or \MSC[2008] code \sep code (2000 is the default)

\end{keyword}

\end{frontmatter}

%% \linenumbers

\section{Introduction} \label{sec:intro}

Systems engineering manages design complexity by modeling with hierarchical, modular architectures. Modular architectures recursively partition complex systems into simpler subsystems which communicate solely through well-defined interfaces. Such models  are  called \emph{architectures} because they represent system structure. SysML \cite{OMG-SysML} is the most popular, general, systems modeling language, though many prefer Capella \cite{Capella} for good reasons.  

The Architecture Analysis \& Design Language (AADL) \cite{AADL-A5506:2022}
was designed specifically to model the electronics and software portions of cyber-physical (embedded) systems. AADL is SAE International standard AS5506D and can be used in conjunction with more general systems engineering languages such as SysML and Capella.

The proposed formalization of AADL RTS in \cite{ISOLA2022} was extensive, but lacked any concept of time. Crucially, in AADL the concept of \emph{event} denotes occurrence at a particular instant of time, and \emph{event data} as transmission/reception of information at a particular instant of time.
This work both extends and simplifies the formalization in \cite{ISOLA2022} adding RTS necessary to support reactive software which receives and sends AADL events and event data using a modal logic defined by a Kripke structure.

Modal logics define many ``worlds'' in which formulas of the logic are evaluated.  Consequently, any particular formula may be true in some worlds and false in others.  In the following, worlds will be distinguished by time (in seconds) as a real number starting at 0, when the system begins functioning, and up to at the present instant, 
$\now$.

% ISOLA 2021 - Add connections to documentation theme
%We believe that appropriately documenting the semantics of the AADL
%RTS is an important step to meeting the challenges above.   While one
%typically thinks of documentation as it applies to a specific model or
%program, in our setting, documentation is needed at the meta-level,
%i.e., for the \emph{definition of the modeling language}.  We argue in
%Section~\ref{sec:documentation} that, with the future of programming
%being increasingly supported by modeling and meta-programming,
%arriving at effective solutions for documentation at different
%meta-levels of the programming framework is important for grounding
%the next generation of programming environments.   Formally specifying
%the semantics of a modeling language is not a new idea.   Rather, our
%contribution is to tackle the challenges of documenting key aspects of
%a modeling language that is being used for rigorous industry model-driven
%development in a manner that can support evolution of
%the modeling language standard and associated ecosystem.    The main
%benefit of the work is to begin to fill a gap that, once closed, can
%provide a much more solid foundation for a community that is already
%invested in rigorous system engineering with AADL.

\subsection{Contributions}
The contributions of this paper are:
\begin{itemize}
\item Add time to the formalization in \cite{ISOLA2022} of the primary notions of thread state and communication
  state within an AADL system.  
%  These notions provide the foundation
%  of conformity assessment by specifying what portions of a executing
%  system's state must be documented for traceability to definitions
%  within the AADL standard and observed for conformance testing based
%  on execution traces.
%\item Develop rule-based formal documentation of the AADL standard's
%  run-time services \emph{including time}, providing a basis for soundness of AADL analysis
%  and verification and the ability to develop conformity assessment
%  for AADL run-time libraries.
\item Identify additional run-time services and support functions
%  needed to fill gaps in the current AADL standard
  necessary for software that issues and responds to events.
\item Generate executable code to perform the additional RTS proposed here.
\end{itemize}

%\subsection{Goals for this Paper}
%Combined together with \cite{ISOLA2022}, o
We aim to propose  a
formal definition of services for AADL RTS \emph{including time} for upcoming major revisions to
the AADL standard.  
We simplified the semantics of \cite{ISOLA2022} to focus on timing behavior, and believe
the semantics described here can be applied to its more expansive, yet time-free semantics.

%Expansion of this formalization can serve as a
%specification for establishing the correctness of AADL analyses and
%code generation.  Our formalization aims to support the \emph{general}
%nature of the AADL RTS  \emph{including time} and to provide a foundation for creating
%\emph{specialized} semantic definitions for particular models of
%computation or platforms.  From an implementation view, this provides
%a means by which implementers that optimize communication pathways or
%thread management can justify executions in their context to be sound
%refinements of the standardized general notions.
%
%It is important that a proposal for aspects of a standardized AADL
%run-time be based on significant implementation experience with
%multiple code generation approaches, multiple languages, and multiple
%target platforms.
%% XXX JHu: Warning, the following identifies the authors
The formalization presented here is informed by
significant experience with HAMR code generation. 
HAMR code generation has been extended to include the timing semantics
described herein, and the additional RTS necessary for software implementing AADL
components using event and event data ports.  

%Whether Ocarina \cite{Gilles-al:AdaEurope2009} can be similarly extended with time
%is uncertain at this time.

The modal logic defined in this paper, to formalize AADL run-time services, embodies the compute-entrypoint/dispatch-status and receive-execute-send para\-digms.  The compute-entrypoint/dispatch-status paradigm allows evaluation of dispatch conditions used by both BA and BLESS.  The read-execute-send para\-digm cleanly separates the semantics of application actions from infrastructure actions.

\subsection{Other Paradigms Supported by AADL}

Of course, as an architecture language, AADL wants to be able to model the systems users have--not necessarily best practice or elegantly formalizable.  

AADL RTS allows invocation of {\tt Receive\_Input} which transfers values from infrastructure to application at any point of computation which breaks the receive-execute-send paradigm.  

AADL supports a port-subprogram paradigm for thread execution by designation of a subprogram to be invoked upon event, or event data arrival individually for each port, as a different paradigm of thread invocation.  HAMR supports the port-subprogram paradigm by generating a Slang \cite{hatcliff2021slang} skeleton which invokes a port's designated subprogram upon event (data) arrival.
However, the port-subprogram paradigm does not allow evaluation of dispatch conditions defined in the BA annex standard, which are also used in BLESS.  %The port-subprogram paradigm breaks  the entrypoint/dispatch-status and receive-execute-send para\-digms.

%AADL allows data components to model persistent state accessible to all components which have data access required features--even though globally-shared variables are shunned--and even defines resource locking and critical section arbitration to try to control concurrent accesses.  BA even defines grammar for locking and critical section RTS which breaks the receive-execute-send paradigm.

Obviously, a single formalism for all AADL run-time services is impossible; they have conflicting assumptions.  Because we are trying to formalize system-level RTS timing
with concision, brevity, and perspicuity, we chose the entrypoint\-/dispatch-status and receive-execute-send paradigms in the definition of RTS semantics.

\subsection{Outline}
Section \ref{sec:background} presents the thread and port state concepts from \cite{ISOLA2022}, while section \ref{sec:differences} describes the conceptual differences in this paper.  \ref{sec:structural-semantics} explains the correspondence between declarative and instance architecture models. Section \ref{sec:temporal} introduces the temporal semantics using modal logic. Section \ref{sec:standardrts} presents excerpts from the AADL standard defining RTS being formalized. Section \ref{sec:threads} discusses thread life cycle. Section \ref{sec:director} discusses the concepts of a Director to coordinate system operation and dispatch threads by invoking thread entrypoints. Section \ref{sec:hamr} details additional RTS implementation specifically for HAMR tool set. Section \ref{sec:conclusion} presents our conclusions.

\section{Background} \label{sec:background}

The AADL standard was created by SAE International to embody systems engineering best practices for design of avionics.
These best practices are applicable to design of all cyber-physical systems--particularly those which are safety critical.  Consequently, the AADL standard is applicable to design of any cyber-physical system.
General systems engineering best practices are collected in the INCOSE Systems Engineering Handbook \cite{INCOSEseh}.

Crucially, AADL enforces (sub-)system boundaries.  System components interact sole through well defined interfaces.  AADL explicitly separates what is visible externally into component types, hiding internals in system implementations.  System implementations can be realized directly, or as subcomponents and connections between them.  Thus systems must be modeled as a containment hierarchy of components, all of which interact solely through well-defined interfaces.

Additionally, component classes are defined (with semantics) that allow systems architecture to be expressed as two, interrelated levels.
A \emph{functional} architecture describes the behavior of the system, mostly realized in software.  A \emph{physical} architecture describes the machine to be manufactured and installed.  Functional components are bound to the physical components which will perform the function.

AADL also defines run-time services (RTS) including communication between functional components and scheduling/dispatch of software implementing functional components needed to conform to AADL's defined semantics.

In the rest of this section we present a brief overview of the formalization of AADL RTS in \cite{ISOLA2022}, to which we add formalization of time.  In AADL, an \emph{event} is an occurrence at a particular moment of time which may be communicated from one component to another.  Events signify something which is true about the system at the moment they are sent.  Therefore, a temporal model of RTS is required to express such truths.

\begin{figure*}[t]
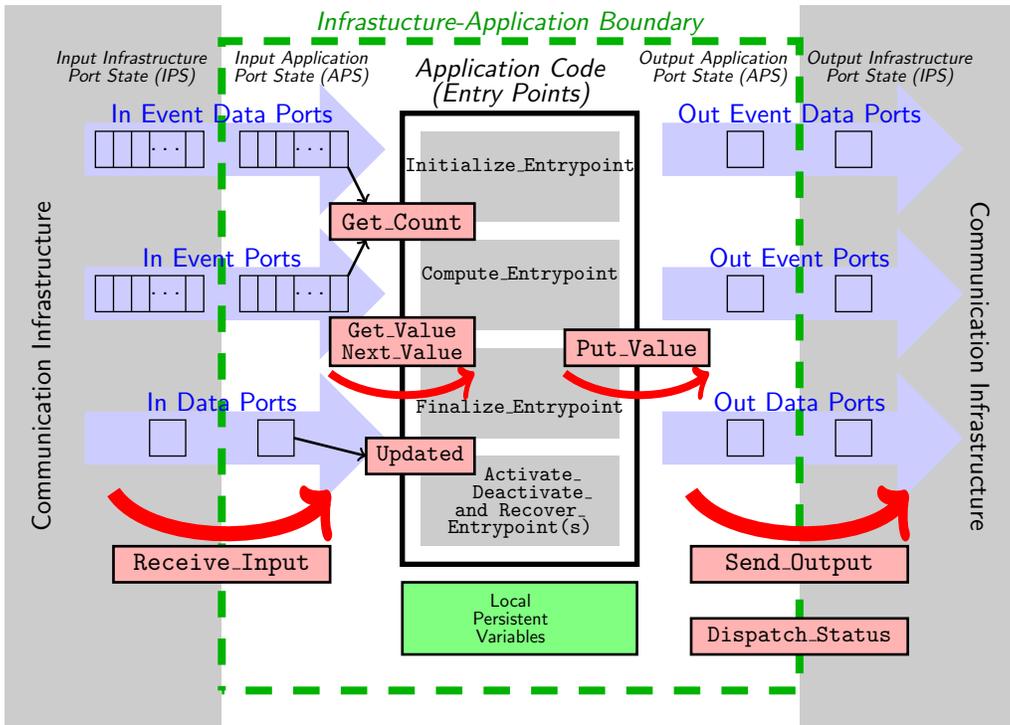

%  \centerline{\includegraphics[width=\textwidth]{concepts.pdf}}
  \threadportconcepts{0.96}
  \vspace{-.4cm}
  \caption{Thread and Port State Concepts (adapted from \cite{ISOLA2022}) }
 \label{fig:state-concepts}
\end{figure*}

Figure \ref{fig:state-concepts} illustrates the relationships between \emph{infrastructure} ports and \emph{application} ports and information provided at thread dispatch, in relation with the thread and port state concepts. An AADL infrastructure provides services to (software) threads in keeping with the AADL philosophy that AADL components need not be concerned with anything other than their explicitly-declared features. Although AADL defines several classes of features, we focus on features most used by threads:  ports.

A port is a gateway through which information is sent or received.  Information may be of three kinds:
\begin{description}
\item[event] an occurrence at a particular time, such as a switch closing
\item[data] a value read at a time of the thread's choosing, such as reading a temperature sensor
\item[event data] a value transmitted at a particular time and queued until a thread is ready to process it
\end{description}

%\begin{figure*}[ht]
%  \centerline{\includegraphics[width=\textwidth]{concepts.pdf}}
%  \vspace{-.4cm}
%  \caption{Thread and Port State Concepts from  \cite{ISOLA2022} }
% \label{fig:state-concepts}
%\end{figure*}

An important contribution of \cite{ISOLA2022} was decomposition of thread ports
into the Infrastructure Port State (IPS) and the Application Port State
(APS), which we adopt. The IPS represents the communication infrastructure's
perspective of the port. The APS represents the thread application
code's perspective of the port.
For input event data ports, the IPS typically would be a queue into
which the middleware would insert arriving values following overflow
policies specified for the port.  For input data ports, the IPS
typically would be a memory block large enough to hold a single value.
For output ports, the IPS represents pending value(s) to be propagated
by the communication infrastructure to connected consumer ports.

%Section \ref{sec:standardrts} quotes AS5506D \cite{AADL-A5506:2022} definitions for the services and entrypoints of Figure \ref{fig:state-concepts}.

Table \ref{tab:rts-entrypoints} summarizes function of RTS and entrypoints of Figure \ref{fig:state-concepts}. 
%Three main types of RTS and entrypoints are discussed \textit{i)} \emph{Infrastructure-Application Boundary},  \textit{ii)} \emph{Invoked by Application},  and \textit{iii)} \emph{Invoked by Director}.}

\begin{table}[htbp]
\centering
\caption{RTS and Entrypoints } \label{tab:rts-entrypoints}
\begin{tabular}{|p{0.35\textwidth}|p{0.6\textwidth}|}
\hline
\multicolumn{2}{|c|}{\textit{Infrastructure-Application Boundary}} \\ \hline
\texttt{Receive\_Input } &  Move values from all input IPS to input APS \\ \hline
\texttt{Send\_Output } & Move values from all output APS to output IPS \\ \hline
\texttt{Dispatch\_Status } &  Information about event (data) arrival which might have caused thread dispatch \\ \hline
\multicolumn{2}{|c|}{\textit{Invoked by Application}} \\ \hline
\texttt{Get\_Value } &  Read an input APS value \\ \hline
\texttt{Next\_Value } & Remove a value from an incoming event data APS queue, allowing \texttt{Get\_Value} to read the next value in the queue \\ \hline
\texttt{Get\_Count } &  The number of elements of an input APS queue \\ \hline
\texttt{Updated } &  Whether the value of in data APS has changed \\ \hline
\texttt{Put\_Value } &  Write a value to out APS \\ \hline
\multicolumn{2}{|c|}{\textit{Invoked by Director}} \\ \hline
\texttt{Initialize\_Entrypoint }  & Code invoked to initialize a thread \\ \hline
\texttt{Compute\_Entrypoint }  & Code invoked by thread dispatch by a Director \\ \hline
\texttt{Finalize\_Entrypoint }  & Code invoked to terminate a thread \\ \hline
\texttt{Activate\_Entrypoint } &  Code invoked when a mode change adds a thread not in the previous mode \\ \hline
\texttt{Deactivate\_Entrypoint } & Code invoked when a mode change removes a thread not in the upcoming mode \\ \hline
\texttt{Recover\_Entrypoint } &  Code invoked to recover from errors or malfunction \\ \hline
\end{tabular}
\end{table}

% \paragraph{Infrastructure-Application Boundary:}
% \begin{description}
% \item[\texttt{Receive\_Input }]  Move values from all input IPS to input APS.
% \item[\texttt{Send\_Output }]  Move values from all output APS to output IPS.
% \item[\texttt{Dispatch\_Status }]  Information about event (data) arrival which might have caused thread dispatch.
% \end{description}

% \paragraph{Invoked by Application:}
% \begin{description}
% \item[\texttt{Get\_Value }]  Read an input APS value.
% \item[\texttt{Next\_Value }]  Remove a value from an incoming event data APS queue, allowing \texttt{Get\_Value} to read the next value in the queue.
% \item[\texttt{Get\_Count }]  The number of elements of an input APS queue.
% \item[\texttt{Updated }]  Whether the value of in data APS has changed.
% \item[\texttt{Put\_Value }]  Write a value to out APS.
% \end{description}

% \paragraph{Invoked by Director:}
% \begin{description}
% \item[\texttt{Initialize\_Entrypoint }]  Code invoked to initialize a thread.
% \item[\texttt{Compute\_Entrypoint }]  Code invoked by thread dispatch by a Director.
% \item[\texttt{Finalize\_Entrypoint }]  Code invoked to terminate a thread.
% \item[\texttt{Activate\_Entrypoint }]  Code invoked whe
n a mode change adds a thread not in the previous mode.
% \item[\texttt{Deactivate\_Entrypoint }]  Code invoked when a mode change removes a thread not in the upcoming mode.
% \item[\texttt{Recover\_Entrypoint }]  Code invoked to recover from errors or malfunction.
% \end{description}

\begin{figure*}[t]
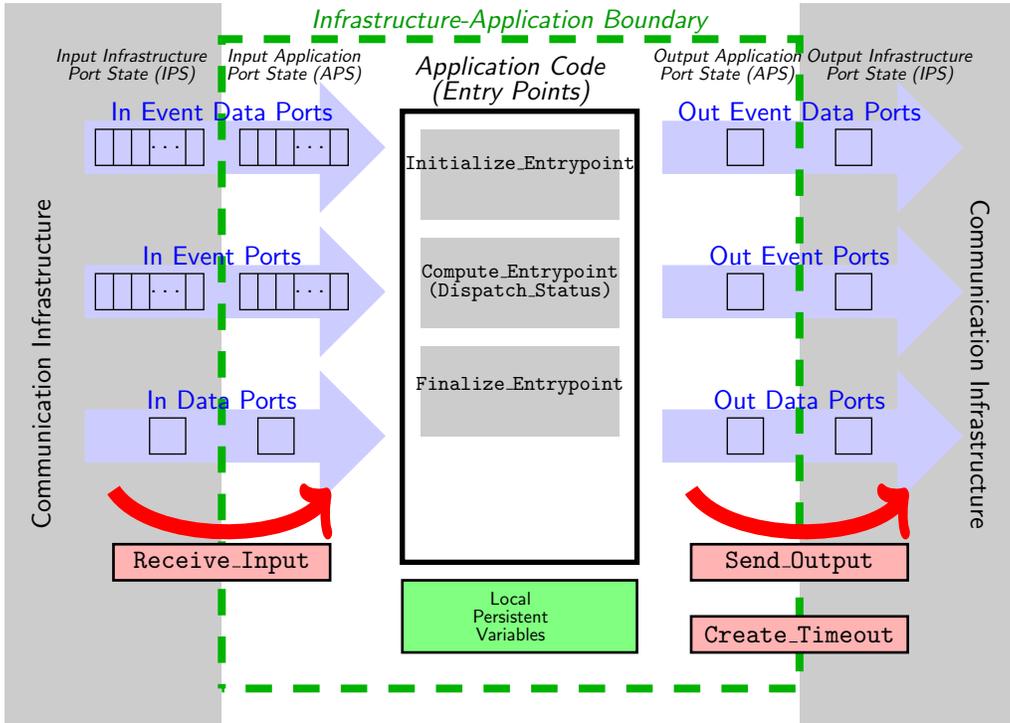

%  \centerline{\includegraphics[width=\textwidth]{concepts.pdf}}
  \ourthreadportconcepts{0.96}
  \vspace{-.4cm}
  \caption{RTS and Entrypoints for Timing Semantics}
 \label{fig:our-state-concepts}
\end{figure*}

In this paper, we consider a subset of RTS and entrypoints of Figure \ref{fig:state-concepts}, and add the additional {\tt Create\_Timeout} RTS in Figure \ref{fig:our-state-concepts}.
{\tt Dispatch\_Status} has been changed from a service, to a parameter of {\tt Compute\_Entrypoint} having a set of active dispatch triggers 
to be evaluated by threads to determine which, if any, dispatch conditions are met.
%which could have caused the Director to invoke {\tt Compute\_Entrypoint}. 

\section{Conceptual Differences}\label{sec:differences}
Our thread and port concepts %depicted in \ref{fig:state-concepts} 
differ from  \cite{ISOLA2022} in the following ways:

\paragraph*{Simultaneous Input and Output:} In \cite{ISOLA2022}, {\tt Receive\_Input} and {\tt Send\_Output} may apply to a subset of ports, and may be performed at any time during execution.  This is a correct interpretation of {\tt Receive\_Input} and {\tt Send\_Output} from the standard.
We presume the read-execute-send paradigm, which uses only a subset of the functionality defined in the standard.   {\tt Receive\_Input} applies to all thread input ports and must be invoked after dispatch and before execution.
Similarly, {\tt Send\_Output} applies to all thread output ports and must be invoked after execution and before suspension.

\paragraph*{Event-like Ports:}
We separate event ports and event data ports instead of using ``event-like" ports.
Event data ports (may) need queues with size larger than one to hold event-data packets from a thread
dispatched more frequently, or from multiple threads connected to the same event data port.
% Event ports need only a queue size of one, because either an event has arrived since the time of the thread's previous suspension, or not.
% That multiple events arrived at an event port is meaningless.

\paragraph*{Output Queues:}
Our output ports hold at most one value.  This greatly simplifies both thread semantics, and infrastructure operation which needs to transmit at most a single value from each out event data port.  No use case for output queue sizes larger than one have been identified.  Should multiple values need to be sent on the same out event port between thread dispatch and suspension, array types can be  used.

\paragraph*{Timeouts:}
Both the BA and BLESS have transition dispatch conditions which include timeout dispatch triggers.
\begin{lstlisting}
T3: sense -[on dispatch timeout (n or p) lrl]-> pace {...};
\end{lstlisting}
Expiration of a timeout may cause thread dispatch similar to arrival of an event or event data.
A timeout RTS is missing from the AADL standard.  Because BA is defined in an AADL annex standard document \cite{AADL-Annex-Vol2:Standard}, and crucial behavior requires timeout dispatch triggers,
a timeout RTS should be added to the standard.

%\paragraph{Frozen Ports:}
%In \cite{ISOLA2022}, an AADL thread may explictly call the RTS to receive
%new inputs at any point in its execution.
%Yet, this would break the atomicity of a dispatch
%execution. We explictly forbid this in our formalization (by ``freezing" input port values) as this would
%introduce unsoundess
%%% JHa: Add the following phrase to clarify what would be unsound.
%in many AADL analyses and contract languages \cite{Cofer-al:NFM12,Larson-al:NFM2013}.
%Furthermore, this capability is barely used in practice.
%Port values are frozen by calling {\tt Receive\_Input} immediately after dispatch, and no-when else.

\paragraph{Dispatch Status:}
\texttt{Dispatch\_Status} is changed to a parameter of \texttt{Compute\-\_Entry\-point} containing a list of ports and timeouts which could have triggered dispatch.
Whether dispatch is accepted is determined by  \texttt{Compute\-\_Entry\-point} given its current state by evaluating dispatch conditions of transitions leaving that state.

\paragraph{Retention of Dispatch Triggers by Dispatch Status Upon Deferred Dispatch:}
In AADL terminology, dispatching a thread refers to the thread
becoming ready for execution from a OS scheduler perspective.  The
thread {\tt Dispatch\-\_Protocol} property selects among several
strategies for determining when a thread should be dispatched.  In
this paper, we consider only {\tt Periodic}, which dispatches a thread
when a certain time interval is passed, {\tt Sporadic}, which
dispatches a thread upon arrival of events and event-data to input ports specified
as {\em dispatch triggers}, and {\tt Timed} which dispatches each period like {\tt periodic}, 
When a thread is dispatched, information
describing the reason for its dispatch accessible via the {\tt Dispatch\-\_Status} parameter to {\tt Compute\_Entrypoint}.  For
example, in a sporadic component, {\tt Dispatch\-\_Status} holds
the set of dispatch triggers (either event arrival or timeout expiration) which might trigger dispatch. 
This may be
used by the component to determine whether to accept the dispatch (or wait
for subsequent dispatch) and to choose reactions accordingly.  
%If dispatch is deferred, \aadlDispatchStatussym\/ retains its dispatch triggers.
%This corresponds
%to evaluation of dispatch conditions used by both BLESS and BA to
%determine which, if any, transition to execute.

If none of the dispatch conditions of transitions leaving the current state are met so no transition is executed, the
dispatch triggers in \aadlDispatchStatussym\/ must be retained, to which additional dispatch
triggers may subsequently added.
If dispatch is accepted, \aadlDispatchStatussym\/ is reset to empty.

For example, suppose a thread having {\tt Dispatch\_Protocol} of {\tt Sporadic} 
has two in event ports, A and B, and currently resides in complete state X. 
State X has a single transition with state X as its source state:

\begin{lstlisting}
X -[ on dispatch A and B ]-> Y {...};
\end{lstlisting}

If an event arrives on port A, but no event arrives on port B, the transition will not be executed.
If later, and event arrives on port B, and the prior event on port A is retained by  \aadlDispatchStatussym\/,
then the transition will be executed. 

\paragraph*{Port Overflow Handling:}  In \cite{ISOLA2022} much attention is devoted to behavior of port {\tt Overflow\_Handling\_Protocol}.
Here we assume that {\tt Port\_Queue\_Size} is sufficiently large to preclude overflow, or that overflow is inconsequential.

\paragraph*{Dequeue Protocol:} Entire input queues are transferred from infrastructure ports to application ports which are then entirely accessible to thread software.
We assume {\tt Dequeue\_Protocol} {\tt => AllItems} rather than the standard default of {\tt OneItem}.

\section{Structural Semantics}
\label{sec:structural-semantics}
The AADL standard includes informal rules that specify how model
elements are organized in terms of containment hierarchy, allowed relationships (e.g., connections,
bindings) with other components, and modeling patterns for system
configuration and deployment. 

A \emph{declarative} model specifies system structure using
components, ports, and connections.  This is the AADL text (or equivalent diagrams) written by system architects.
An \emph{instance} model, representing a specific product, is generated from a declarative model.

AADL is powerful enough to describe many products within the same declarative model which may represent design alternatives, or entire product lines.  
Most analysis tools use instance models because they are inherently simpler than declarative models, and represent a specific product. 

% The intent of
% this section is not to exhaustively capture all declarative and instance model
% information, but rather the information that influences the execution
% semantics presented in the following sections.

Consequently, we will restrict our attention to declarative models corresponding to a single product without loss of generality.

\subsection{Declarative Model}

Declarative architectural models represent components as component \-
types having corresponding component implementations.  Component types express everything externally visible of the component (features and properties) while component implementations express their internals, which may have subcomponents and connections between them, or behavior definitions using BLESS or BA annex subclauses.  
\begin{lstlisting}
system TypeID
  features
    . . .
  properties
    . . .
end TypeID;
\end{lstlisting}
\begin{lstlisting}
system implementation TypeID.ImplementationID
  subcomponents
    . . .
  connections
    . . .
  properties
    . . .
end TypeID.ImplementationID;
\end{lstlisting}

An AADL component can have one of several categories, including software (i.e., \texttt{thread}, \texttt{process}, \texttt{subprogram}), hardware (i.e., \texttt{processor}, \texttt{mem\-ory}, \texttt{bus}, \texttt{device}), and general (i.e., \texttt{system} and \texttt{abstract}).

In this document, we focus on a subset of component categories used to define \emph{functional} architectures: $\mathit{cat}_C \in \{\threadsym, \processsym, \systemsym,$ $ \devicesym\}$ to which AADL RTS apply.
A \emph{physical} architecture defines the structure of the machine built to which components of a functional architecture are bound (i.e. threads to processors).

Properties $\propsym$ specify relevant characteristics of component types and implementations for
%the detailed 
design and implementation such as dispatch protocol and thread period.
% descriptions from an external perspective.  

A \emph{feature} describes an interface of a component through which control and data may be provided to or required from other components.
\begin{definition}
A feature $f \in \featsym$ is a tuple $(f_{id}, \mathit{cat}_F, d, \mathit{type}, \propsym)$ such that:

$f_{id}$: is the unique feature identifier,

$\mathit{cat}_F$: is the category of the feature $\mathit{cat}_F \in \{\datasym, \eventsym, \eventdatasym\}$,

$d$: is the direction of the feature, where $d \in \{\mathit{in}, \mathit{out}\}$,

$\mathit{type}$: is a data component defining the type of $f$ and

$\propsym$: is the set of properties associated with this feature.

\end{definition}
The AADL standard defines several categories of features. We focus on a subset of categories: data ports, event ports and event data ports.\footnote{
In AADL, other categories of features denote subprogram parameters, abstract features, or access to resources. They do not directly participate in the semantics of threads and are omitted in this paper.}. 
Given that we focus on ports as a relevant subset of feature categories, we use the terms \emph{port} and \emph{feature} interchangeably.
The direction $d$ of a feature corresponds to either \texttt{in} or \texttt{out}. A feature $\mathit{type}$ corresponds to a data component instance.
Feature properties $\propsym$ are configuration parameters that specify relevant characteristics of the feature and its implementation.

AADL component types define everything visible externally of the component, in particular, its features.
Hiding internal information is crucial to modular architectures.

\begin{definition} 
A component type $\mathit{Ctype}$ is a tuple $(c_{ty}, \mathit{cat}_C, \featsym, \mathit{Prop})$ s.t.

$c_{ty}$: the unique component type identifier,

$\mathit{cat}_C$: the component category $\in \{${\tt thread process system device}$\}$,

$\featsym$: the set of features,

$\propsym$: the set of properties associated with this component type.

\end{definition}

AADL component implementations define the internal of components.  A component implementation specifies which component type it implements. A component type may have more than one implementation, each with a unique implementation identifier.

Composite components have subcomponents and connections between them.  Composite components manage the hierarchical decomposition of complex components into simpler ones.  

AADL allows subcomponents to be names of either component types or component implementations.  This allows a single AADL declarative model to represent product-line families by defining multiple implementations for the same type.  An instance model can be generated for a particular product by selecting which implementations should be instantiated.

\begin{definition}
A component implementation $\mathit{Cimp}$ is a tuple $(c_{ty}, c_{im}, S, X,$ $\propsym, A)$ s.t.

$c_{ty}$: the unique component type identifier,

$c_{im}$: the unique component implementation identifier,

$S$: the set of subcomponents, 

$X$: the set of ordered pairs of features representing directional connections between
features of $S$, and also $\featsym$, the features of the component type being implemented,

$\propsym$: the set of properties associated with this component implementation, and

$A$: an annex subclause defining component behavior.

\end{definition}

Generally, an implementation may have either subcomponent and connections, or an annex subclause defining behavior, but not both.

A \emph{top level} system implementation contains subcomponents, but is not itself a subcomponent.  Its subcomponents may themselves have subcomponents forming a containment hierarchy. % which we call $\modsym$. The set of all components of $\modsym$ we refer to as $\compsym$.

%Given that all component identifiers are unique at each layer of the instance model, navigation of the hierarchy is performed through top-down access of component identifiers starting from the root component $r$ and traversing down the containment (subcomponent) hierarchy, e.g., $r.\mathit{childId}$.

%These can be viewed as rules of conformance between the
%described components and their implementation. A detailed summary of
%relevant component properties addressed in this document may be found
%in our related tech report \cite{AADL-RTS-FORM:TR}.

\subsection{Instance Model}\label{sec:instance}

An \emph{instance model}, $\modsym$, is created from a top-level system implementation, recursively creating instances of subcomponents by merging their type with their implementation and flattening the containment hierarchy.  %Most AADL analysis tools apply to instance models.

For simplicity, we assume that all components in the declarative model without subcomponents are threads, so the instance model, $\modsym$ consists of threads and connections between them.  

AADL device components, though hardware, behave similarly to threads in that they can originate and respond to events, and may have annex subclauses defining their behavior.
Here we model device components as threads.

\begin{definition}
An architecture instance model $\modsym$ is a pair ($T_M$, $X_M$) s.t.

$T_M$: is a set of thread components created by merging a thread type, $Ctype$, with its implementation,
$Cimp$, and

$X_M$: is a set of connections (ordered pairs) between features of threads in $T_M$.  Therefore, $(p_1,p_2) \in X_M$ iff there is a connection from port $p_1$ to port $p_2$ in the instance model.
\end{definition}

%We assume that $\modsym$ is completely instantiated and bound. %A set of contained property associations can reflect property values specific to individual instances of components, ports, connections, and accesses (clause 13.1(4)). We assume that all specified subcomponents, ports, and subprograms are complete with regard to their specification.

% In this document,
%We refer to the global set of unique thread component identifiers as $\compids$. 
We use the meta-variable $\compv \in \compids$ to range over thread identifiers defined by replacing the period with an underscore in the name of thread implementations.
Port (feature) identifiers % $f_{id}$ 
can also be accessed globally in
$\modsym$ similar to thread identifiers. We use the meta-variable
$p \in \portids$ to range over port identifiers, sometimes adding a subscript as with the definition above.

To access the
elements of $\modsym$, a number of helper methods
and predicates are defined. 
%For more detail on these predicates, see
%our companion tech report \cite{AADL-RTS-FORM:TR}.
%% WARNING blind review violation
% \ifnoblindreview\cite{AADL-RTS-FORM:TR}\else\emph{suppressed due to blind review}.\fi.

\begin{definition}
Instance Model Helper Methods
\begin{itemize}
\item $\modelportkind{\portv}$ returns $y \in \{\datasym, \eventsym, \eventdatasym\}$
\item $\modelisdataport{\portv}$ returns $\truesym\/$ iff $\modelportkind{\portv}\,=\,\datasym$
%\item $\modeliseventport{\portv}$ returns $\truesym\/$ iff $\modelportkind{\portv}$ returns $y \in \{\eventsym\}$
%\item $\modeliseventlikeport{\portv}$ returns $\truesym\/$ iff $\modelportkind{\portv}$ returns $y \in \{\eventsym, \eventdatasym\}$
\item $\modeloutports{\compv}$ returns all outgoing ports for thread $\compv$
\item $\modelinports{\compv}$ returns all incoming ports for thread $\compv$
%\item $\modelcompoutdataports{\compv}$ returns all $\outdataportssym$ for component $\compv$
%\item $\modelcompouteventports{\compv}$ returns all $\outeventportssym$ for component $\compv$
%\item $\modelcompouteventlikeports{\compv}$ returns all $\outeventlikeportssym$ for component $\compv$

\item $\modelportdirection{\portv}$ returns $y \in \{\insym, \outsym\}$
\item $\modelisinport{\portv}$ returns $\truesym\/$ iff $\modelportdirection{\portv} \,=\, \insym$
\item $\modelisoutport{\portv}$ returns $\truesym\/$ iff $\modelportdirection{\portv} \,=\,\outsym$
\item $\modelconndest{\compv}$ returns the set of all connection destination ports from $p$

%\item $\modeldequeuepolicy{p}$ returns $y \in \{\allitemssym, \oneitemsym\}$
\item $\modeldispatchprotocol{p}$ returns $y \in \{\periodicsym, \sporadicsym\}$
%\item $\modeloverflowhandlingprotocol{p}$ returns $y \in \{\dropoldestsym, \dropnewestsym, \errorsym\}$
\end{itemize}
\end{definition}

%temporal-semantics.tex

\section{Temporal Semantics}\label{sec:temporal}

These temporal semantics model \emph{ideal} time rather than \emph{actual} time.  Time is modeled as a real number, whereas any deployment would discretize time in clocking of synchronous logic.  A correct deployment would behave within some tolerance of its ideal model.

Operations which are supposed to be fast occur with ``negligible" delay which is modeled using infinitesimal duration $\delta$.  Moving values from IPS to APS upon execution of 
{\tt Receive\_Input} should happen with negligible delay, $\delta$.

Of course, deployed systems would perform {\tt Receive\_Input} taking some time greater than infinitesimal.  Whether this actual time would be brief enough to be ``negligible" will depend on tolerances determined by the designers.  For example, implantable pacemakers and cardio-defibrillators act in millisecond increments for which execution of {\tt Receive\_Input} in less than 50 microseconds would be negligible.

A \emph{modal} logic defines symbols whose meaning differs depending in which ``world" the symbol is evaluated.
A \emph{temporal} logic is a modal logic for which each world corresponds to an instant or interval of time.
A good synopsis of other temporal logics used can be found in \cite{Konur-2013}, which surveys 18 of them.

\begin{definition}
A dynamic architecture model, $\mathfrak{M}$ is defined as a tuple
 
 \[
 \mathfrak{M} \equiv \langle \mathbf{T}, \prec,  \modsym, D, \valsym, \mathbf{I}, O \rangle
 \] such that
 
$\mathbf{T}$: is a set of time instants, 

$\prec$: is a temporal precedence relation, 

$\modsym$: is a (static) instance model,

$D$: is a set of dynamic variables,

$\valsym$: is the domain of values for variables, 

$\mathbf{I}(d,\tau)$: is an interpretation for $d \in D$ at time $\tau \in \mathbf{T}$, and

$O$: is a set of arithmetic, relational, and logical operators.
\end{definition}

\subsection{Time: $\mathbf{T}$}\label{sec:time}

The many ``worlds" of this modal logic $ \mathfrak{M}$ corresponding to instants of time $\tau\in\mathbf{T}$, beginning $(\tau =0)$ when the system starts operation, until the present instant $(\tau =\mathsf{\msc{now}})$.  The instants of time are real numbers expressing duration since starting system operation, in seconds.
 
\begin{definition}\label{def:df-bl.time} Define domain of time $\mathbf{T}$
from commencement of operation, $0$ to the present instant, $\mathsf{\msc{now}}$:
%\begin{align}
\[
%\mm{\vdash}
 \mathbf{T} = \{ \tau~\vert~( \tau \in \mathbb{R} \wedge 0 \le
    \tau \wedge \tau \le \mathsf{\msc{now}} )
    \} %\label{eq:df-bl.time}\tag{df-bl.time}
%\end{align}
\]

where $\mathbb{R}$ is the set of real numbers, and $\tau$ represents the duration in seconds since beginning operation.  

\end{definition}

Thus, a timestamp service is needed to supply $\tau\in\mathbf{T}$ corresponding to the present instant, $\mathsf{\msc{now}}$. 

The additional \aadlTimeStampsym ~RTS approximates the model timestamp with floating-point numbers: 

\begin{lstlisting}
subprogram Time_Stamp
features
  ts: out parameter Time_t;
end Time_Stamp;  
\end{lstlisting}

\noindent where \lstinline{Time_t} is equivalent to \lstinline{Base_Types::Float} (see \S\ref{sec:valuedomain} in its implementation.

Because every platform has some form of timestamp, \aadlTimeStampsym ~RTS can be provided by format conversion.
%\footnote{This format conversion may be challenging as our experience providing \aadlTimeStampsym  ~with HAMR revealed.}

\subsection{Temporal Precedence: $\prec$}\label{sec:precedence}

%All instant models of time define a relation of \bemph{temporal precedence}, $\prec$, a.k.a ``before".
%Models of time can choose many different properties for $\prec$.  The properties of ATL temporal precedence are marked with *; other temporal logics choose different temporal precedence properties.
%
%Define temporal precedence; a 
A time is before ($\prec$) another when its numeric value is smaller.

\begin{definition}\label{def:df-bl.before} Temporal Precedence (before)
\[
%\begin{align}
( \tau_1 \prec \tau_2 \leftrightarrow \tau_1 < \tau_2
    )  %\label{eq:df-bl.before}\tag{df-bl.before}
%\end{align}
\]
\end{definition}

It's also useful to define ``before or coincident".  

\begin{definition}\label{def:df-bl.beforeeq} Temporal Precedence (before or
coincident)
%\begin{align}
\[
\mm{\vdash} ( \tau_1 \preccurlyeq \tau_2 \leftrightarrow ( \tau_1 <
    \tau_2 \vee \tau_1 = \tau_2 )
    )  %\label{eq:df-bl.beforeeq}\tag{df-bl.beforeeq}
%\end{align}
\]
\end{definition}

\subsection{Dynamic Variables: $D$}\label{sec:variables}

Dynamic variables, $D$, are the Infrastructure Port State and Application Port State associated with each port in the system together with local persistent values within each thread.

For each $p \in \portids$, let $\ips_p \in D$ be its Infrastructure Port State, and $\aps_p \in D$ be its Application Port State.

For each $\compv \in \compids$, let $\variables_t \subset D$ be the local persistent variables of thread $t$.

All values have an inherent timestamp indicating when they were created.  For value $v\in D$, $ts(v)$ is the timestamp of the value's creation.

\subsection{Value Domain: $\valsym$}\label{sec:valuedomain}

Data ports and event data ports specify their datatypes with the name of an AADL data component type.
The Data Modeling annex \cite{AADL-Data-Model} to the AADL standard defines a property set, \aadlDataModelsym, 
and a package of commonly-used data component types, \aadlBaseTypessym, defined with properties from \aadlDataModelsym.

The predeclared property set \aadlDataModelsym~allows definition of common atomic types (boolean, string, integer, floating-point, enumeration), and constructors (array, record, variant).  Consequently, values of such constructed types always use the same number of bytes which can be calculated at compile time.

The value domain  $\valsym$ is equivalent to types definable with \aadlDataModelsym~annex standard document.

We adopt the representation in \cite{ISOLA2022} for
IPS and APS for event, event data, and data ports: a queue
$\queuev$ is a bounded sequence of values.   The bound is specified by the 
\aadlQueueSizesym ~property of the port which defaults to 1.

 $\queuemake{v}$ denotes a
queue with a single data value $v$;
 $\queuemake{v_2~v_1}$ denotes a
queue with a data value $v_1$ followed by $v_2$; and,

% e.g., as for a data port, and
$\queueempty$ denotes an empty queue.
%$\queuesnoc{\queuev'}{\valuev}$ represents a queue with $v$ being
%the first item to be dequeued (head) and $\queuev'$ being the rest of
%the queue.  This representation is specialized for data ports and event ports. 
For a data port, the queue
size is always one and enqueueing overwrites the previous value.  For
event ports, queues hold 0 or more values, and $\eventpresent$ denotes the
presence of an event in a queue (e.g., $\queuedevent$ is a queue
holding a single event). 
The maximum size of queues for a particular event and event data port is
statically configurable by AADL port property, \aadlQueueSizesym, which must ensure the port queue cannot overflow.  

Defensive design would detect deviations of operational behavior from intended behavior (errors) such as port queue overflow during operation and handle them to restore intended behavior as closely as possible.
However, this is outside the scope of formal modeling of correct behavior, here.

\subsection{The Interpretation:  $\mathbf{I}$}\label{sec:interpretation}
 
The interpretation $\mathbf{I}$ defines the meaning of dynamic symbols $d \in D$ at time $\tau \in \mathbf{T}$, in the value domain $\mathbf{I}(d,\tau) \in \valsym$.  

 Some of the dynamic symbols represent inputs to the system, sensors like a temperature measurement or a switch closing.  These are monitored variables.  Other dynamic symbols represent outputs of the system, actuators like a motor or a LED.  These are controlled variables.

All other dynamic symbols are internal to the system:  persistent variables and states of threads, or port queues.  

%Execution of BLESS programs \emph{causes} the values of variables given by $\mathbf{I}_{\tau}$.  
Execution, together with dynamic inputs to the machine being controlled, creates a model $\mathfrak{M}$.

%When we say that a program is \bemph{correct}, that means that all possible models  $\mathfrak{M}$ having interpretation $\mathbf{I}_{\tau}$ will conform to the program's specification. 
 
 Sometimes, it is more convenient to say $d@\tau$ than $\mathbf{I}(d,\tau)$:
 \begin{definition}
 \[d@\tau \equiv \mathbf{I}(d,\tau)\]
 \end{definition}
 
 Later, the use of $@$ will be expanded to evaluation of predicates or expressions at a particular time.

\subsection{Operators: $O$}\label{sec:operators}

All operators are constant having conventional meanings as formally defined in Metamath database \texttt{set.mm} \cite{set.mm}.
 
\subsection{Infinitesimal Duration $\delta$}

Often, a tiny bit of time is useful for modeling.

\begin{definition}
Infinitesimal Duration $\delta$ is

\[ \delta \equiv \lim_{\tau \to 0} \tau
\] 
\end{definition}

Therefore, $(\tau \prec \tau + \delta)$.  Using $\tau'$ as shorthand for $\tau + \delta$, then $(\tau \prec \tau')$.

%standardrts.tex

\section{Excerpts from the AADL Standard with Operational Semantics}\label{sec:standardrts}

The AADL standard AS5506D \cite{AADL-A5506:2022} defines Runtime Support For Ports in Section 8.3.5.
We quote definitions for {\tt Receive\_Input}, {\tt Send\_Output}, and {\tt Dispatch\_Status} below, with quoted section number in square brackets, followed by discussion of how they fit into the operational concepts of the previous section.

%In many cases, the subprograms allow inclusion of detail deemed unnecessary or irrelevant to our formalization.
\subsection{Receive\_Input}
From AS5506D:
\begin{quotation}
[A.9 (5)] A {\tt Receive\_Input} runtime service allows the source text of a thread to explicitly request port input on its incoming ports to be frozen and made accessible through the port variables.  Any previous content of the port variable is overwritten, i.e., any previous queue content not processed by {\tt Next\_Value} calls is discarded.  The {\tt Receive\_Input} service takes a parameter that specifies for which ports the input is frozen.  Newly arriving data may be queued, but does not affect the input that thread has access to (see Section 9.1).  {\tt Receive\_Input} is a non-blocking service. 
\end{quotation}
\begin{lstlisting}
subprogram Receive_Input
features
  InputPorts: in parameter 
    <implementation-dependent port list>;
    -- List of ports whose input is frozen
end Receive_Input;
\end{lstlisting}    

To adhere to the \textit{receive-execute-send}  paradigm, {\tt Re}\-{\tt ceive\_Input()} copies \emph{all} of the input IPS ports to input APS ports, thus the {\tt InputPorts} parameter
must include all incoming ports.  {\tt Receive\_Input()} is called immediately after the dispatch has been accepted by the thread, and prior to computation.  For incoming data ports, 
{\tt Receive\_Input()} overwrites the value of the APS port.  For incoming event data ports, {\tt Receive\_Input()} adds values of the IPS port queue to the APS port queue.
% {\color{blue}
%\begin{lstlisting}
%subprogram Receive_Input
%end Receive_Input;
%\end{lstlisting}    }

\subsection{Send\_Output} 
From AS5506D:
\begin{quotation}
[A.9 (3)] A {\tt Send\_Output} runtime service allows the source text of a thread to explicitly cause events, event data, or data to be transmitted through outgoing ports to receiver ports.  The {\tt Send\_Output} service takes a port list parameter that specifies for which ports the transmission is initiated. The send on all ports is considered to occur logically simultaneously.  {\tt Send\_Output} is a non-blocking service.  An exception is raised if the send fails with exception codes indicating the failing port and type of failure. 
\end{quotation}
\begin{lstlisting}
subprogram Send_Output
features
  OutputPorts: in parameter 
    <implementation-dependent port list>;
  -- List of ports whose output is transferred
  SendException: out event data; 
  -- exception if send fails to complete
end Send_Output;
\end{lstlisting}    

The parameter features of subprogram {\tt Send\_Output} accept a list of output ports, and return a send exception if one occurs.
Conformance to the read-compute-write paradigm sends outputs together after computation is complete, so the {\tt OutputPorts} parameter includes all outgoing ports.
%Our definition of infrastructure never raised an exception when executing {\tt Send\_Output}, so 
%Similarly the {\tt SendException} parameter is unnecessary.

 {\tt Receive\_Input} and  {\tt Send\_Output} are formalized in \S\ref{sec:receivesend}.

\subsection{Dispatch\_Status}
The {\tt Dispatch\_Status} runtime service has three mentions in the AADL standard.

\noindent From AS5506D:
\begin{quotation}
In the latter case, the source text code sequence can determine the context of the execution through a  {\tt Dispatch\_Status} runtime service call. [5.4.1 (23)]
\end{quotation}

\begin{quotation}
This function then invokes an implementer-provided  {\tt Dispatch\-\_Sta\-tus} runtime service call to identify the context of the call and to branch to the appropriate code sequence. [5.4.8 (93)]
\end{quotation}

\begin{quotation}
The thread can determine the port that caused the error by calling the standard  {\tt Dispatch\_Status} runtime service. [8.3.3 (25)]
\end{quotation}

Rather than an explicit RTS subprogram, we choose {\tt Dispatch\_Status} to be a list of identifiers of active dispatch triggers as parameter of {\tt Compute\-\_Entry\-point}.

\subsection{Entrypoints}

The AADL standard defines thread lifecycle [5.4.1] and dispatching [5.4.2] with hybrid automata.  These automata describe invoking \emph{entrypoints} for various purposes.
Each entrypoint is a subprogram.

%Entrypoints are specified using properties applied to threads.
\noindent From AS5506D:
\begin{quotation}
Entrypoints for thread execution can be specified in three ways: by identifying source text name, by identifying a subprogram classifier representing the source text, or by a call sequence.  [5.4]
\end{quotation}

\begin{quotation}
An {\tt Initialize\_Entrypoint} (enter the state performing \\thread initialization in Figure 5) is executed during system initialization and allows threads to perform application specific initialization, such as ensuring the correct initial value of its out and in out ports. A thread that has halted may be re-initialized. [5.4.1 (24)]
\end{quotation}

%\begin{quotation}
%The {\tt Activate\_Entrypoint} (enter the state performing thread activation in Figure 5)  and {\tt Deactivate\_Entrypoint}  (enter the state performing thread activation in Figure 5) are executed during mode transitions and allow threads to take user-specified actions to save and restore application state for continued execution between mode switches.  These entrypoints may be used to reinitialize application state due to a mode transition.  Activate entrypoints can also ensure that out and in out ports contain correct values for operation in the new mode.  [5.4.1 (25)]
%\end{quotation}

\begin{quotation}
The {\tt Compute\_Entrypoint} (enter state performing thread computation in Figure 5) represents the code sequence to be executed on every thread dispatch.  Each provides subprogram access feature represents a separate compute entrypoint of the thread. Remote subprogram calls are thread dispatches to the respective entrypoint.  Event ports and event data ports can have port specific compute entrypoints to be executed when the corresponding event or event data dispatches a thread.  [5.4.1 (26)]
\end{quotation}

%\begin{quotation}
%A {\tt Recover\_Entrypoint} (enter the state executing recovery in Figure 7) is executed when a fault in the execution of a thread requires recovery activity to continue execution. This entrypoint allows the thread to perform fault recovery actions (for a detailed description see Section 5.4.4).  [5.4.1 (27)]
%\end{quotation}

\begin{quotation}
A {\tt Finalize\_Entrypoint} (enter the state performing thread finalize in Figure 5) is executed when a thread is asked to terminate as part of a process unload or process stop.  [5.4.1 (28)]
\end{quotation}

%\begin{quotation}
%If no value is specified for any of the entrypoints, then there is no invocation at all.  [5.4.1 (29)]
%\end{quotation}
%
%The last statement is ominous, but need not be.  HAMR will generate a skeleton for each entrypoint not specified by a thread property with names  {\tt Initialize\_Entrypoint}, etc.

In our formalization, {\tt Compute\_Entrypoint} is a subprogram having three parameters:  context, dispatch status (set of active dispatch triggers), returning whether the dispatch was accepted.

\begin{lstlisting}
subprogram Compute_Entrypoint
features
  api : in parameter <implementation defined context>;
  Dispatch_Status : in parameter 
    <implementation defined list of active dispatch triggers>;
  dispatched : out parameter Base_Types::Boolean;  
    --whether dispatch was accepted
end Compute_Entrypoint;
\end{lstlisting}  

In our formalization, {\tt Initialize\_Entrypoint} and  {\tt Finalize\_Entrypoint} are subprograms having a single context parameter.

\begin{lstlisting}
subprogram Initialize_Entrypoint
features
  api : in parameter <implementation defined context>;
end Initialize_Entrypoint;
\end{lstlisting}
\begin{lstlisting}
subprogram Finalize_Entrypoint
features
  api : in parameter <implementation defined context>;
end Finalize_Entrypoint;
\end{lstlisting} 

Thread entrypoints are formalized in \S\ref{sec:threadentry}.

\section{Threads}
\label{sec:threads}

This section formalizes thread operation
described in Section~5.4 of the AADL standard, AS5506D.  
%Thread life cycle is depicted in Figure \ref{fig:figure5} (taken from the standard).
Only a subset of defined AADL entry points and RTS are formalized--those most critical to operation and timing.

% Section~\ref{sec:director} addresses the Director's aggregation of
% thread states into system states and RTS that are used in the
% coordination of thread actions with platform scheduling and communication.

\subsection{Thread Life Cycle}

%\begin{figure*}
%  \vspace{-0.5cm}
% \centerline{\includegraphics[width=\textwidth]{AADL-RTS-documentation-thread-states.png}}
% %\vspace{-0.5\baselineskip}
% \vspace{-.4cm}
% \caption{Old Thread Operational Life-Cycle}
% \vspace{-.4cm}
%\end{figure*}

\begin{figure*}
  \centerline{\includegraphics[width=\columnwidth]{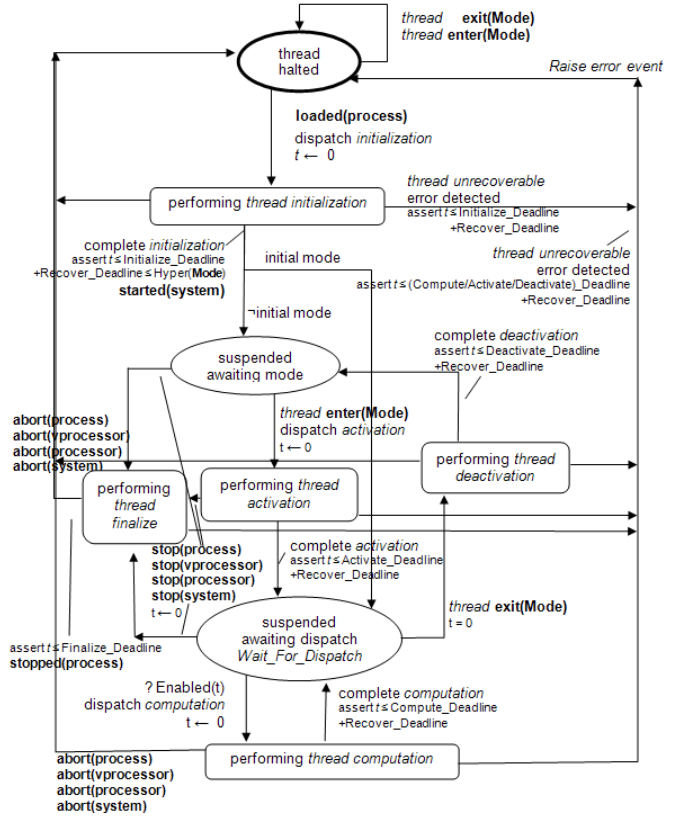}}
  \vspace{-.4cm}
  \caption{AADL Standard Thread Timed Automaton}
 \label{fig:figure5}
\end{figure*}

\begin{figure*}
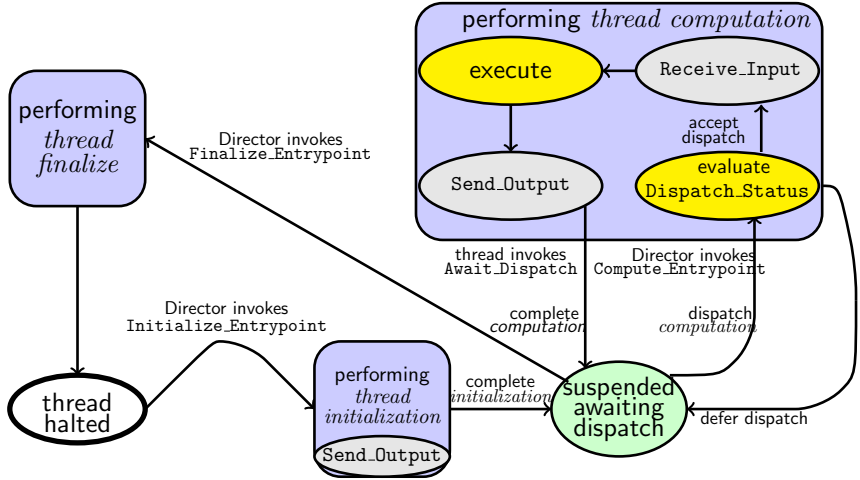
%[ht]
%  \vspace{-0.5cm}
\threadlifecycle{0.9}
 %\vspace{-0.5\baselineskip}
 \vspace{-.4cm}
 \caption{Thread Operational Life-Cycle}
 \vspace{-.4cm}
\label{fig:thread-operational-life-cycle}
\end{figure*}

The AADL standard defines a thread's operational life cycle as a timed automaton depicted in Figure \ref{fig:figure5}.
Like  \cite{ISOLA2022} we simplify the standard's timed automata in Figure \ref{fig:figure5} by removing mode
switches, error recovery, preemption, and termination shown in Figure~\ref{fig:thread-operational-life-cycle}.  

\textbf{thread halted}:~~
After power-on the thread waits in the ``thread halted" state.  

\textbf{performing {\it thread initialization}}:~~
The Director then invokes {\tt Initial\-ize\_Entrypoint} so the thread can initialize its persistent variables, and put values on all of its out data ports with {\tt Send\_Output}.  Together with the {\tt Move} communication action performed by the Director, this ensures that all input IPS data ports have legitimate values.
When initialization is complete the thread is suspended, awaiting dispatch. 

\textbf{suspended awaiting dispatch}:~~
Later, the Director invokes {\tt Compute\-\_Entrypoint} when a dispatch trigger of the thread occurs.

\textbf{performing {\tt thread computation}}:~~
The thread evaluates the dispatch triggers in the {\tt Dispatch\_}\-{\tt Status} parameter of {\tt Compute\_Entrypoint} to determine whether to accept or defer the dispatch.
If dispatch is accepted, the thread performs {\tt Receive\_Input}, executes, performs {\tt Send\_Output} and is suspended when computation is complete. Active dispatch triggers are cleared. 
If dispatch is deferred, the thread is immediately suspended, and active dispatch triggers retained.

\textbf{performing {\it thread finalize}}:~~
To terminate thread execution, the Director invokes {\tt Finalize\_Entrypoint}.

\subsection{Thread State}

The state
of each thread $t$ is formalized as a 6-element tuple
\[
\compcont{\iniportsc}{\inportsc}{\outportsc}{\outiportsc}{\cvarsc}{\dsv}.
\]

$\iniportsc$ (IPS) maps each input port identifier of the
thread $t$, $\modelinports{\compv}$, to a queue holding buffered messages.
For data ports, the queue is always non-empty and has a size of one.

$\inportsc$ (APS) maps each input port identifier of the
thread $t$, $\modelinports{\compv}$, to a queue representing the application code's view of
frozen port values during its execution.

The entire queues of $\iniportsc$ are transferred to $\inportsc$ together upon {\tt Receive\_Input}.

$\outportsc$  (APS) maps each output port identifier for the thread $t$,  $\modeloutports{\compv}$,
to a queue of size one, representing the application code's view of
output port.  

% \item
$\outiportsc$ (IPS) maps each output port identifier of the
thread $t$,  $\modeloutports{\compv}$, to a queue of size one.

The queues of $\outportsc$ are transferred to $\outiportsc$ upon {\tt Send\_Output}.

For any of these port state structures, the notation
$\mapupdate{\inportsc}{p}{\queuev}$ denotes
an updated port state structure that is like the original $\inportsc$
except that port $\portv$ now maps to the queue $\queuev$.

$\cvarsc$ represents a thread's local
state, e.g., a mapping from variables to values,
that may be accessed by the application code during entrypoint
execution.  Since manipulation of local variables is independent of the
semantics of the AADL RTS, we omit a more detailed formalization.

$\dsv$ holds information, accessible during entry point execution,
that provides information about the thread's dispatch status--particularly the set of active dispatch triggers.

\subsection{\aadlReceiveInputsym\/ and \aadlSendOutputsym\/ RTS}\label{sec:receivesend}

\aadlReceiveInputsym\/  transfers input messages (data) from IPS to APS.
Conversely, \aadlSendOutputsym\/  transfers output messages (data) from APS to IPS.

Our modeling, here, of \aadlReceiveInputsym~differs from \cite{ISOLA2022} in that all data from all infrastructure port queues is transferred to all application port queues together.  This both upholds the \textit{receive-execute-send} paradigm and simplifies the semantics.

The rules in Figure~\ref{fig:receive-input} formalize the \aadlReceiveInputsym\/
RTS.  In Figure~\ref{fig:receive-input}, elements of a thread tuple without apostrophe are evaluated at time $\tau$, i.e.
$\inportsc\equiv\inportsc@\tau$ and  elements of a thread tuple with apostrophe are evaluated at time $\tau'=\tau+\delta$, i.e.
$\inportsc{'}\equiv\inportsc@\tau{'}\equiv\inportsc@(\tau+\delta)$.

\begin{figure}[h]
%\vspace{-0.6cm}
\begin{frameit}
{\scriptsize%\vspace{-0.4cm}
\[
  %%   R e c e i v e    I n p u t  --   D a t a
  \tinferlabt{\aadlrule{Receive\_Input for thread t at time $\tau$}}
     {\evalrecinp{\compcont{\iniportsc}{\inportsc}{\outportsc}{\outiportsc}{\cvarsc}{\dsv}}
                   {\compv}{p}
                   {\compcont{{\iniportsc}'}{{\inportsc}'}{\outportsc}{\outiportsc}{\cvarsc}{\dsv}}}
     { %%  s t a t i c     m o d e l      c o n d i t i o n s
     \forall p \in \modelinports{\compv} 
           \left| \begin{array}%[b]
                      {r@{\;}c@{\;}l}
                      {\iniportsc}' & = &\iniportsc ~~ $if$~\modelisdataport{\portv}
                     \\
                      {\iniportsc}' & = & \mapupdate{\iniportsc}{p}{\queueempty}~~$otherwise$
                     \\
                     {\inportsc}' & = & \mapupdate{\inportsc}{p}{\maplookup{\iniportsc}{p}}
                    \end{array}\right.
       }
\]
}
\end{frameit}
%\vspace{-0.4cm}
\caption{\texttt{Receive\_Input} RTS Rule}
\label{fig:receive-input}
%\vspace{-0.2cm}
\end{figure}

% The rules are stated for a single port, but extend to a list
%of ports (matching the informal subprogram interface from the
%standard) in a straight-forward way.

The {\tt Receive\_Input} rule has relational structure:
\[
\evalrecinp{\compcont{\iniportsc}{\inportsc}{\outportsc}{\outiportsc}{\cvarsc}{\dsv}}
                   {\compv}{\portv}
                   {\compcont{{\iniportsc}'}{{\inportsc}'}{\outportsc}{\outiportsc}{\cvarsc}{\dsv}}
\]
The $\evalrecinprelsym{\compv}$ relational symbol indicates
that thread $\compv$ (typically component infrastructure code) is invoking the
\aadlReceiveInputsym\/ service on the all ports of $\compv$.
As a side-effect of the invocation, the state
$\compcont{\iniportsc}{\inportsc}{\outportsc}{\outiportsc}{\cvarsc}{\dsv}$
of thread $\compv$ is transformed to a new state
$\compcont{{\iniportsc}'}{{\inportsc}'}{\outportsc}{\outiportsc}{\cvarsc}{\dsv}$.
% Since the purpose of the service is to transfer data/messages from the
% thread's port infrastructure into the view of the application code,
% the rules only modify the infrastructure port state $\iniportsc$
% (e.g., dequeueing for event-like ports) and application port state
% $\inportsc$ (e.g., updating the thread's input port variables).
The rule conclusions indicate that only the $\iniportsc$ and
$\inportsc$ portions of state are modified, e.g.,  by moving/copying values from
the IPS into the APS. 
% In source code, the call to the service would
%take the form $\mbox{RecInP}(\portv)$ (here \mbox{RecInP} abbreviates
%the full name of the service).
%% e.g., as indicate in the AADL standard
%% and associated Code Generation Annex).
The other elements of the rule are implicit in the code execution context,
e.g., $\compv$ is the currently active thread with the associated
thread states.  
The clause above the rule horizontal bar represents the antecedent
to the rule.  
Both of the rules has an implicit side condition that its state
elements $\iniportsc$, etc. are compatible with the model's port
declarations for $\compv$.

In Figure~\ref{fig:send-output}, the rule for \aadlSendOutputsym\/
specifies that values are removed from the output ports of APS, $\outportsc$, into the output ports of IPS, $\outiportsc{'}$.
\begin{figure}[h]
%\vspace{-0.6cm}
\begin{frameit}
{\scriptsize%\vspace{-0.4cm}
\[
    %%   S e n d     O u t p u t   --
  \tinferlabt{\aadlrule{Send\_Output for thread t at time $\tau$}}
     {\evalsendoutp{\compcont{\iniportsc}{\inportsc}{\outportsc}{\outiportsc}{\cvarsc}{\dsv}}
                   {\compv}{p}
                   {\compcont{\iniportsc}{\inportsc}{{\outportsc}'}{{\outiportsc}'}{\cvarsc}{\dsv}}}
                   { %%  s t a t i c     m o d e l      c o n d i t i o n s
                    \forall p \in \modeloutports{\compv} 
                    \left|  \begin{array}%[b]
                           {r@{\;}c@{\;}l}
                          {\outportsc}' & = & \mapupdate{\outportsc}{p}{\queueempty}
                          \\
                          {\outiportsc}' & = & \mapupdate{\outiportsc}{p}{\outportsc(p)}
                          \end{array} \right.
                                          }
\]
}
%\vspace{-0.4cm}
\end{frameit}
%\vspace{-0.4cm}
\caption{\texttt{Send\_Output} RTS Rule}
\label{fig:send-output}
%\vspace{-0.6cm}
\end{figure}
%
%There is only a single rule because the action is the same regardless
%the type of port.

Invoking {\tt Send\_Output} on thread $t$ at time $\tau$ changes only the application and infrastructure ports which get new values (having transferred application to infrastructure) at time $\tau'=\tau+\delta$.

\subsection{Timeout RTS}\label{sec:timeoutrts}

Both BA and BLESS have timeout dispatch triggers which must be supplied as RTS.
A timeout dispatch trigger has a list of event (data) ports which can (re-)start the timeout, and a duration.
Whenever an event occurs on one of listed ports at the duration previously, and no events on any of the listed ports since,
the timeout causes thread's \texttt{Compute\_Entrypoint} to be invoked, and the identifier of the timeout included in the dispatch triggers
returned by \aadlDispatchStatussym.

\begin{lstlisting}
subprogram Create_Timeout
  features 
    resetPorts:  in parameter 
      <implementation-dependent port list>
    duration: in parameter <implementation-dependent time>
    triggerID: out parameter 
      <implementation-dependent port identifier>
end Create_Timeout;
\end{lstlisting}  

The \texttt{resetPorts} must be event or event data ports, either in or out, of the thread to received the dispatch trigger.
The \texttt{duration} must be a time-valued AADL property or property constant, or a time-valued data port of the thread.
The value returned by \texttt{triggerID} must be the port identifier ($p \in \portids$) included in \aadlDispatchStatussym\/ to indicate timeout expiration.

\begin{definition} A timeout dispatch trigger $to$ is included in the dispatch triggers returned by \aadlDispatchStatussym~at time $\tau$ when:

\[ to@\tau \iff ( \exists p \in r~|~ p@(\tau - d)~) \wedge ( \forall q \in r~ |~ ( \lnot \exists \tau_2 \in [ \tau - d ,, \tau ] ~|~ q@\tau_2 ) ~ )
\]
such that

$r$: is the \texttt{resetPorts} parameter of the \texttt{Create\_Timeout} invocation,

$d$: is the \texttt{duration} parameter of the \texttt{Create\_Timeout} invocation,

$to$: is the port identifier, $(to \in \portids)$ returned as \texttt{triggerID} parameter of the \texttt{Create\_Timeout} invocation, and

$[ \tau - d ,, \tau ]$: is the open interval after $\tau - d$ and before $\tau$.

\end{definition}

A timeout dispatch trigger occurs when one of the reset ports sent or received an event $d$ before $\tau$, and no events were sent or received by any of the reset ports.

\subsection{Thread Entrypoints}\label{sec:threadentry}

%In Figure~\ref{fig:thread-entrypoints}, there are two rules for each
%entrypoint category: the {\em AppCode} rules reflects the execution of
%``user's code'', while the {\em Infrastructure} rules captures the code
%structure that AADL code generation tools would typically use for
%enforcing AADL's \emph{Read; Compute; Write}
%paradigm.   The distinct {\em AppCode} rules allow us to parameterize
%% reflect the fact that our formalizationis parameterized on
%% the semantics of the
%the semantics with the user code (thus, the use of
%the distinct $\Rightarrow$ symbol), i.e., our
%framework semantics for AADL RTS, entry point concepts,
%etc., is orthogonal to the application-specific semantics of the
%entry point user code.

The \aadlInitEP\/ application code rule for a thread $\compv$
must uphold the AADL standard's statements that
% indicates that one of its key purposes is to
(a) no input port values are to be read during the initialization
phase (access to $\inportsc$ is not provided to the code),
(b) a thread's \aadlInitEP\/ should set initial values for
\emph{all} of $\compv$'s output data ports (i.e., after execution completes, all data port queues in
${\outportsc}'$ are required to have size of 1, and (c)
sending initial values on event (data) ports is forbidden.

\begin{figure}[h]
%\vspace{-0.4cm}
\begin{frameit}
{\scriptsize%\vspace{-0.4cm}
\[
  %%   I n i t i a l i z e    E n t r y    P o i n t   (App Code)   --
  \tinferlabt{\aadlrule{Initialize\_Entrypoint for thread t}}
     {\Rightarrow^{\mbox{\tiny Initialize\_Entrypoint}}_t\langle{\outportsc}' , {{\cvarsc}'}\rangle }
                   { %% o u t    p o r t    c o n s t r a i n t s
                     \begin{array}[t]{c}
                        $initialize$(\cvarsc) ;
   \\                    
                      \forall p \in \modeloutports{\compv} 
                    \left|  \begin{array}{l}
                   {\outportsc}'(p)\,=\,\queuemake{default(p)}~~ $if$~\modelisdataport{\portv}
                     \\
                      {\outportsc}'(p)\,=\,\queuemake{\cdot} ~~ $otherwise$
                    \end{array} \right.
                       \\   ;~~  $Send\_Output$_t
                    \end{array}
                   }
\]
}
\end{frameit}
%\vspace{-0.4cm}
\caption{Initialize Entrypoint Semantics}
\label{fig:initialize}
%\vspace{-0.6cm}
\end{figure}

\aadlInitEP ~must be completed by every thread, and all of the infrastructure initialized, before
commencement of operation at time 0.  Thus \aadlInitEP ~is not in model $\mathfrak{M}$;
it occurs \emph{before} the beginning of $\mathbf{T}$, establishing the correct interpretation $\mathbf{I}$
for time 0.  Establishing correct values for $\outportsc$ is one of many responsibilities of \aadlInitEP

The \aadlComputeEP\/ user code rule
indicates that the user code computes first evaluates its \texttt{Dispatch\_Status}, $\dsv$,
to determine whether it should do anything at all.
If so, it invokes \texttt{Receive\_Input}, computes
a function
from the input APS $\inportsc$, local variable
state, $\cvarsc$, and dispatch status, $\dsv$, to output APS values ${\outportsc}'$ and (updated) local
variable state, ${\cvarsc}'$, and finally invokes {\tt Send\_Output}.  Semicolons indicate required ordering of operations.
Recall that {\tt Receive\_Input}, if invoked, moves infrastructure input port queues, $\iniportsc$, to application input port queues $\inportsc$.

\begin{figure}[h]
%\vspace{-0.6cm}
\begin{frameit}
{\scriptsize%\vspace{-0.4cm}
\[
  %%   C o m p u t e    E n t r y    P o i n t   (App Code)   --
  \tinferlabt{\aadlrule{Compute\_Entrypoint for thread t}}
%           {\evaltacc{\inportsc}{\outportsc}{\cvarsc}{\dsv}{{\outportsc}'}{\cvarsc'}}
          {{\compcont{\iniportsc}{\inportsc}{\outportsc}{\outiportsc}{\cvarsc}{\dsv}}\Rightarrow^{\mbox{\tiny Compute\_Entrypoint}}_t{\compcont{{\iniportsc}'}{{\inportsc}'}{{\outportsc}'}{{\outiportsc}'}{{\cvarsc}'}{\dsv}}}           
            { 
            evaluate(\dsv) \rightarrow
             \left\{ \begin{array}{l}
			{Receive\_Input}_t~;
			\\
			\langle{\inportsc}{\cvarsc}{\dsv}\rangle {\Rightarrow^{\mbox{\tiny Compute}}_t \langle{\outportsc}'}{\cvarsc'}\rangle ;
			\\
			{Send\_Output}_t
             \end{array} \right\}
             } 
\]
}
\end{frameit}
%\vspace{-0.4cm}
\caption{Compute Entrypoint Semantics}
\label{fig:compute}
%\vspace{-0.6cm}
\end{figure}

The {\tt Finalize\_Entrypoint} user code rule does nothing!  It allows freeing of memory, or other going-away actions, but does not affect thread variables, state, or port-queues.

\begin{figure}[h]
%\vspace{-0.6cm}
\begin{frameit}
{\scriptsize%\vspace{-0.4cm}
\[
  %%  F i n a l i z e   E n t r y    P o i n t   (App Code)   --
  \tinferlabt{\aadlrule{Finalize\_Entrypoint for thread t}}
%           {\evaltacc{\inportsc}{\outportsc}{\cvarsc}{\dsv}{{\outportsc}'}{\cvarsc'}}
          {\Rightarrow^{\mbox{\tiny Finalize\_Entrypoint}}_t }           
            { ~~~~~~~~~~~~~~~~~~~~~~~~~~~~~~~~~~~~~~~~~~~~~~~~~~~~~~~~
%            evaluate(\dsv) \rightarrow
%             \left\{ \begin{array}[b]{l}
%			{Receive\_Input}_t~;
%			\\
%			(\inportsc,\cvarsc,\dsv) {\Rightarrow^{\mbox{\tiny Compute}}_t ({\outportsc}'}{\cvarsc'}) ;
%			\\
%			{Send\_Output}_t
%             \end{array} \right\}
             } 
\]
}
\end{frameit}
%\vspace{-0.4cm}
\caption{Finalize Entrypoint Semantics}
\label{fig:finalize}
%\vspace{-0.6cm}
\end{figure}

%Overall, these rules enhance the AADL standard by clarifying the specific
%portions of thread state that entry points may read or write and
%indicating other semantic properties.
%The infrastructure rules illustrate how future versions of the
%standard may present code pattern options for enforcing important
%semantic properties of AADL.

\section{Director}
\label{sec:director}

In \cite{ISOLA2022}, \emph{Director} was introduced to invoke thread entrypoints, run timers for timeouts, and transmit events and data from output 
infrastructure ports, to input infrastructure ports.  
Director's operations are specified to provide latitude for scheduling and communication implementations.

\begin{definition}
The $Director$ is defined as a tuple
 
\[
 Director~\equiv~\langle Phs, Thrs, Schs, Comms \rangle
\] such that
 
$Phs$: is the current phase of the Director, \\ $Phs \in \{off, initialize, move, dispatch, compute, finalize\}$, 

$Thrs$: is the combined state of all threads, 

$Schs$: is the schedulability of each state t, \\ $Schs(t) \in \{WaitingForDispatch,Initializing,Computing,Halted\}$,

$Comms$: is the combine Infrastructure Port State (IPS).

\end{definition}

\begin{figure*}[t]
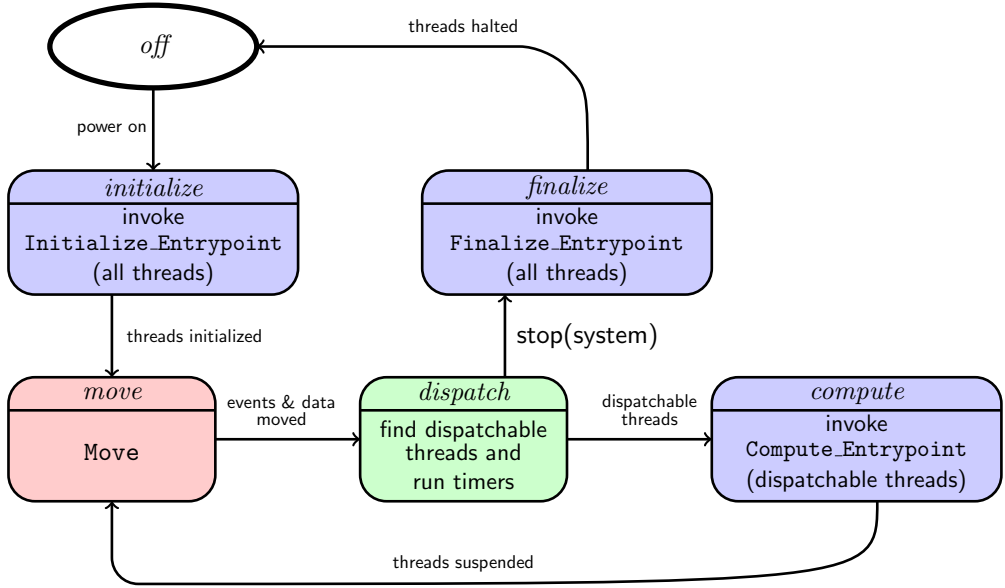

%  \centerline{\includegraphics[width=\textwidth]{concepts.pdf}}
  \directorcycle{1.1}
  \vspace{-.4cm}
  \caption{Director Life Cycle}
 \label{fig:director-operation}
\end{figure*}

Director's operations are depicted in Figure \ref{fig:director-operation} as phases.  Let the current phase be $Phs$.
Initially, the Director phase is $Phs=off$.  
Each of the operations in Figure \ref{fig:director-operation} are described below.

\subsection{\emph{off}}

When $Phs=off$ the system is unpowered.  When the system is turned on, hardware is powered in a specific order together with their clock trees.
If power-on self-test of hardware passes, Director transitions to $Phs=initialize$.

\subsection{\emph{initialize}}

Power-on causes the Director phase to transition to $Phs=initialize$ during which  Director invokes {\tt Initialize\_Entrpoint} for each thread $\compv \in \compids$  which establishes values for all 
output data infrastructure ports, and may also output event and event data ports.

After power on, the Director phase is $Phs=initialize$.
According to Figure \ref{fig:director-initialize}, Director invokes \texttt{Initial}\-\texttt{ize\_Entrypoint} for each thread $t \in T_M$.
When all threads are initialized, Director transitions to $Phs=move$.

\begin{figure}[h]
\begin{frameit}
{\scriptsize
\[
%%    \arraycolsep=2pt
%   \begin{array}{l}
 \tinferlabt{\aadlrule{Initialize System}}
          %% r u l e   c o n c l u s i o n
         {
         (\sysphasev=\sysphaseinit,{\syscommv}) \rightarrow (\sysphasev=\sysphasecompute,{\syscommv}')
          }
           %% a n t e c e d e n t s
           {
             \begin{array}[b]{c}
           \forall \compv \in \compids~ | ~\mathsf{Initialize\_Entrypoint}_t
              \end{array}
   }
%\end{array}
\]
}
\end{frameit}
\caption{Director Initialize}
\label{fig:director-initialize}
\end{figure}

When all threads have completed initialization the Director phase transitions to $Phs=Move$, during which Director performs {\tt Move} to transmit from output infrastructure ports $\outiportsc$ to input infrastructure ports $\iniportsc$ for all $\compv \in \compids$.

After Director has initialized all threads and performed the first {\tt Move}, the interpretation  $\mathbf{I}(d,\tau) \in \valsym$ is established
for $\tau=0$ and all $d \in D$.
The system begins operation.

Note:  Director and its phases are \emph{outside} of the dynamic architecture model, $\mathfrak{M}$ defined in Section \ref{sec:temporal}.  The set of time instants $\mathbf{T}$ does not even start until initialization and the first move are completed.  Thereafter, Director's actions are recorded in $\mathfrak{M}$, but is not part of $\mathfrak{M}$. 

\subsection{\emph{move}}

Figure~\ref{fig:comm-substrate} presents the \texttt{Move} action of the Communication system (middleware).
The \texttt{Move} action is very simple.  For every non-empty infrastructure out port, $(\maplookup{\outiportsc}{p} \: = \: \queuemake{q}) $, copy its value, $\queuemake{q}$ to all infrastructure in ports, $\maplookup{\iniportsc}{p_2}$, for which port $p$ connects to port $p_2$, $(p,p_2) \in X_M$.
Thus, a single out port can have its value broadcast to multiple in ports.  
Similarly, in event (data) ports can receive values from multiple out ports.
In that case, each value is added to the queue in non-deterministic order.

%NOT SURE HOW TO REPRESENT ADDING A VALUE TO A QUEUE.

\begin{figure}[h]
\begin{frameit}
{\scriptsize
\[
%%    \arraycolsep=2pt
%\begin{array}{l}
  %%   C o m m     O u t p u t  
  \tinferlabt{\aadlrule{Move}}
  {\cmove{\sysstate{\sysphasev}{\systhreadsv}{\sysschedv}{\syscommv}}
                   %\compcont{\iniportsc}{\inportsc}{\outportsc}{\outiportsc}{\cvarsc}{\dsv}
            %       {\compv}{\portv}
                   {\sysstate{\sysphasev}{\systhreadsv}{\sysschedv}{\syscommv'}}}
                   { %%  s t a t i c     m o d e l      c o n d i t i o n s
                  
                    \begin{array}{l}
                       \forall \compv \in \compids ~\forall \portv \in \modeloutports{\compv}                   
                    \\
                       %~~~~
                       (\maplookup{\outiportsc}{p} \: = \: \queuemake{q}) \rightarrow
                       \left\{ \begin{array}{l}
                       \forall p_2 s.t. (p,p_2) \in X_M |  {\iniportsc}' \, = \, \mapupdate{\iniportsc}{p_2}{\queuemake{q}}~;
                        \\
                        {\outiportsc}' \, = \, \mapupdate{\outiportsc}{\portv}{\queuemake{}}
                      
                       \end{array} \right\}
                    \end{array}
                   }
 %\end{array}  
\]
}
\end{frameit}
\caption{Communication Service (Move)}
\label{fig:comm-substrate}
\end{figure}

\subsection{\emph{dispatch}}

When $\sysphasev=\sysphasecompute$ the Director alternates between dispatching threads (when appropriate)
and invoking Communication service \texttt{Move} for message (data) transport.
Only threads which are waiting for dispatch are considered for dispatch.
Dispatch status, $\dsv'$, is computed for each dispatchable thread.

\texttt{Compute\_Dispatch\_Status} is determined differently for periodic and sporadic threads.  For periodic threads, $\dsv' = \dstimetriggeredsym$, when the current timestamp, $\now$, is a multiple of its \texttt{Period} property.
For sporadic threads, $\dsv' = \dseventtriggeredsym(\portlist)$ if any of the infrastructure port queues, $\iniportsc$, for events or event data are not empty.
Otherwise, the dispatch status is not enabled, $\dsv' = \dsnotenabled$.

\begin{figure}[h]
\begin{frameit}
{\scriptsize
\[
   \begin{array}{l}
      \tinferlabt{\aadlrule{Dispatch Threads}}
          %% r u l e   c o n c l u s i o n
         {
           \evalDirDispatchThread{\sysstate{\sysphasev}{\systhreadsv}{\sysschedv}{\syscommv}}
                                                     {\compv}
                                                     {\sysstate{\sysphasev}{\systhreadsv'}{\sysschedv'}{\syscommv'}}
          }
           %% a n t e c e d e n t s
            {
             \begin{array}[b]{c}
                {\sysphasev}\,=\,\sysphasecompute
                \\
               \forall \compv \in \compids~ | ~ (\maplookup{\sysschedv}{\compv}\,=\,\sysschedwait) \rightarrow
               \\ ~~~
               \left\{  \begin{array}{l}
                 {\dsv'} := \mathsf{Compute\_Dispatch\_Status}(\now,\iniportsc) ~;
                   \\                                                         
                   ( \dsv' \ne \emptyset ) \rightarrow \mathsf{Compute\_Entrypoint}_t(\dsv')
                 \end{array} \right\} 
                 \\
               ; ~~\mathsf{Move}
             \end{array}
     }
\end{array}
\]
}
\end{frameit}
\caption{Director Dispatch}
\label{fig:director-dispatch}
\end{figure}

The Director will repetitively \emph{dispatch}-\emph{compute}-\emph{move} until the system is stopped by \textbf{stop(system)}.

\subsection{\emph{compute}}

When director is \emph{compute}, {\tt Compute\_Entrypoint}  is invoked for each dispatchable thread. The formal semantics for {\tt Compute\_Entrypoint} are defined Figure \ref{fig:compute}.

Finally, after every \texttt{Compute\_Entrypoint} has completed, the \texttt{Move} service is invoked to transfer 
out infrastructure queues, $\outiportsc$, to in infrastructure queues, $\iniportsc$.

\subsection{\emph{finalize}}

Upon \textbf{stop(system)}\footnote{
AADL also defines \textbf{stop} for smaller portions of the system, \textbf{stop(process)}, \textbf{stop(vprocessor)}, \textbf{stop(processor)}, 
which would halt a subset of all threads, but not included in this formalization for simplicity.
} Director invokes {\tt Finalize\_Entrypoint} for each thread  $\compv \in \compids$.
The formal semantics for {\tt Finalize\_Entrypoint} are defined Figure \ref{fig:finalize}.
Everything after  \textbf{stop(system)} is also outside of $\mathfrak{M}$; operation has ceased.

After all threads are halted, Director transitions to $Phs=\off$.

%hamr-implementation.tex

\section{Additional RTS Implementation by HAMR}\label{sec:hamr}

\underline{H}igh-\underline{A}ssurance \underline{M}odeling and \underline{R}apid Engineering for Embedded Systems (HAMR) \cite{DBLP:conf/isola/HatcliffBRC21} transforms AADL instance models into a software skeleton.  This skeleton provides AADL RTS leaving the behavior code to be either programmed manually, or automatically-generated by another tool.

HAMR is a plugin to the Open-Source AADL Tool Environment (OSATE) \cite{osate} which is itself a plugin to Eclipse.  HAMR generates code starting with a selected top-level system implementation.  The selected system implementation is transformed into an instance model.  From the instance model the skeleton is generated in an IntelliJ project, which is best viewed with the IVE customization of IntelliJ--part of KSU's Sireum tool environment.  The skeleton consists on an IVE project having templates for each thread in the instance model.  The templates, which are to be edited (filled in) by a programmer are written in Sireum's Slang language which is a high-assurance adaptation of Scala to write programs to control safety-critical machines. HAMR already provides a timestamp service, so implementing {\tt time\_stamp} is just format conversion. It also provides the {\tt HamrCodegenPluginProvider} Eclipse extension points with which plugins to HAMR insert automatically-generated Slang text into the skeleton.  HAMR-provided extension points are summarized in Table \ref{fig:hamr-ext}.

%\begin{figure}[htbp]
%\begin{center}

%\begin{center}
%\begin{tabular}{|l{4.55cm}|l{8cm}|}
%\hline
%BehaviorProviderPlugin & insert generated text for Compute\_Entrypoint \\ \hline
%EntryPointProviderPlugin & insert generated text into Infrastructure and Director \\ \hline
%DatatypeProviderPlugin & insert generated text into AADL data components  \\ \hline 
%PlatformProviderPlugin & create and initialize any other needed objects \\ \hline
%\end{tabular}
%\end{center}
%\end{table}%

%\caption{ HAMR Plugins}
%\label{fig:hamr-ext}
%\end{center}
%\end{figure}

\lstset{language=slang}

\begin{table}[htbp]
\centering
\caption{HAMR Plugins} \label{fig:hamr-ext}
\begin{tabular}{|p{0.35\textwidth}|p{0.6\textwidth}|}
\hline
BehaviorProviderPlugin & insert generated text for Compute\_Entrypoint \\ \hline
EntryPointProviderPlugin & insert generated text into Infrastructure and Director \\ \hline
DatatypeProviderPlugin & insert generated text into AADL data components  \\ \hline 
PlatformProviderPlugin & create and initialize any other needed objects \\ \hline
\end{tabular}
\end{table}

The BLESS plugin to OSATE/Eclipse communicates with HAMR by supplying an implementation of each HAMR plugin in Figure \ref{fig:hamr-ext} making four connections to the Eclipse extension point {\tt HamrCodegenPluginProvider}.

The additional RTS of timeouts and dispatch deferment are injected by providing a refinement of EntryPointProviderPlugin.  

Threads may defer dispatch by returning false to {\tt Compute\_Entrypoint} invocation.  If dispatch is deferred, the set of dispatch triggers which have occurred since the time of previous suspension (a.k.a. {\tt Dispatch\_Status}) is retained.  Retained dispatch triggers are added to fresh dispatch triggers to make {\tt Dispatch\_Status} for the next invocation of {\tt Compute\_Entrypoint}.  If dispatch is accepted, all dispatch triggers are cleared.

The {\tt Create\_Timeout} service may only be served during initialization.  Code to do so is injected by the BLESS EntryPointProviderPlugin.  This also adds an event port for each timeout to both IPS and APS which map port names (numbers) to port variables.  Thereafter, each timer is reset with the occurrence of an event of one of the ports listed in {\tt Create\_Timeout} by the infrastructure, and will invoke {\tt Compute\_Entrypoint} of its thread when the timeout duration expired.

The listing which follows was generated by HAMR (in Slang, a dialect of Scala), with help from the BLESS refinement of EntryPointProviderPlugin for a thread having a single timeout in the dispatch conditions of its transitions.

\begin{lstlisting}
def compute(): Unit = {
    operational_api.logDebug("VVI_i_pp_t_Bridge.compute()")  
    val EventTriggered(receivedEvents) = 
      Art.dispatchStatus(VVI_i_pp_t_BridgeId)
    event_set = event_set ++ receivedEvents  
    Art.receiveInput(eventInPortIds, dataInPortIds)
    val dispatched : B = { component.Compute_Entrypoint
        (operational_api, event_set) }
    if (dispatched) { event_set = Set.empty[Art.PortId] }
    if ( Art.observeOutPortVariable(n_Id).nonEmpty | 
        Art.observeOutPortVariable(p_Id).nonEmpty )
        { ArtTimer.scheduleTrait("timeout_n_p_lrl", T, 
            timeout_n_p_lrl_Duration, 
            Callback_timeout_n_p_lrl(this)) }
    Art.sendOutput(eventOutPortIds, dataOutPortIds)
}
\end{lstlisting}

Variable \texttt{event\_set} retains dispatch triggers when dispatch is deferred.  
  Fresh dispatch triggers detected by the infrastructure, \texttt{received\-Events}, are added to \texttt{event\_set}, then passed to 
{\tt Compute\_Entrypoint} as its \texttt{Dispatch\_Sta\-tus} parameter.

\texttt{Receive\_Input} is implemented by  \texttt{Art.receiveInput}; \texttt{Send\_Output} is implemented by  \texttt{Art.sendOutput}.
Note:  both are invoked by infrastructure--not application code--to enforce the \emph{receive-execute-send} paradigm.

Flag \texttt{dispatched} becomes true when \texttt{Compute\_Entrypoint} accepts dispatch, to empty \texttt{event\_set}.

Between \texttt{Compute\_Entrypoint} and  \texttt{Send\_Output} occurrence of events \\which could reset the timer are checked with
\texttt{Art.observeOutPortVariable( \_Id).nonEmpty}.  If so, a timer is scheduled, \texttt{ArtTimer.scheduleTrait}, with timeout duration (which may be variable), and supply a call-back method to be invoked upon timer expiration.

\begin{lstlisting}
  @datatype class Callback_timeout_n_p_lrl(ep : EntryPoints) 
      extends TimerCallback
    {
    override def callback(): Unit = 
      { ep.timeout_n_p_lrl_expires( mc = this ) }
    }
\end{lstlisting}

When a timer expires and invokes the \texttt{callback()} method, an event is added to \texttt{event\_set} indicating the timeout's dispatch trigger occurred.  Then \texttt{compute()} is called (see above).

\begin{lstlisting}
    def timeout_n_p_lrl_expires(mc: Callback_timeout_n_p_lrl): 
      Unit =
      {
      operational_api.logDebug("timeout_n_p_lrl expires")
      event_set = event_set + timeout_n_p_lrl_ID
      compute()
      }
\end{lstlisting}

In Figure \ref{fig:director-operation}, Director invokes invokes \texttt{Compute\_Entrypoint} for all threads.  By semantics of Figure \ref{fig:compute}, \texttt{Compute\_Entrypoint} invokes \texttt{Receive\_Input} and \texttt{Send\_Output} by thread code.  HAMR-generated code moves invocation of \texttt{Receive\_Input} and \texttt{Send\_Output} to infrastructure code, \texttt{compute()}.

Executable byte code generated by later phases of HAMR has been demonstrated with Arduino UNO boards running a Firmata emulation which talks to a JVM running the byte code.
HAMR is multi-targeted to generate executable code for multiple processor architectures including STM32, and real-time operating systems like OpenRTOS or separation kernels like seL4.

\section{Conclusion}
\label{sec:conclusion}

This paper presents an extension to the formal semantics of AADL RTS specified in \cite{ISOLA2022} to include time. Some of the port operations are also simplified to make the temporal semantics more concise and comprehensible to enforce the \emph{receive-execute-send} paradigm.  A modal logic based on Kripke structure is defined to establish ``worlds'' representing instants of system operation using @ as a modal operator to access other worlds. The AADL RTS are defined in terms of this modal (temporal) logic to describe the progression during execution of variable values for both threads and RTS infrastructure.  Additional RTS not included in the AADL standard, but necessary for state-machine implementation of threads are defined. Director semantics, adapted from \cite{ISOLA2022}, are also defined in terms of the modal logic. A summary of the additional AADL RTS implemented by HAMR (an AADL multi-platform code generation toolset) is also presented. HAMR, augmented by BLESS plugins, generates executable byte code that demonstrates the use of the additional RTS functions, including timestamp, timeout, and deferred dispatch.

%It is hoped that the temporal semantics introduced here will be considered for inclusion when formal semantics for AADL RTS are discussed for standardization.

% \oomit{
% The formal semantics presented here extend the formal semantics of \cite{ISOLA2022} to include time.  Some simplifications of port operations of \cite{ISOLA2022}  made the temporal semantics shorter and clearer by enforcing the \emph{receive-execute-send} paradigm.  The richer port operations defined in \cite{ISOLA2022} may violate the \emph{receive-execute-send}  paradigm.  In those cases, these temporal semantics will be inapplicable
% A modal logic was defined with a Kripke structure to define the ``worlds" in which application and infrastructure variables may have different values in different worlds corresponding to instants of system operation.
% AADL RTS were defined in terms of the modal (temporal) logic, defining the progression of variable values during execution.
% Director (adapted from \cite{ISOLA2022}) semantics were defined in terms of the modal (temporal) logic,
% Implementation of the additional AADL RTS by HAMR was briefly summarized.
% HAMR augmented by BLESS plugins generates executable byte code demonstrating use of the additional RTS (timestamp, timeout, and deferred dispatch).
% Hopefully, the temporal semantics defined here will be included when formal semantics for AADL RTS are considered for standardization.
% }

 \bibliographystyle{elsarticle-num} 
 \bibliography{biblio}

%% else use the following coding to input the bibitems directly in the
%% TeX file.

%\begin{thebibliography}{00}

%% \bibitem[Author(year)]{label}
%% Text of bibliographic item

%\bibitem[ ()]{}

%\end{thebibliography}
\end{document}